%% file: QuantumDragon.tex
\documentclass[showpacs,showkeys,nofootinbib,amsfonts,amssymb,floatfix
,twocolumn,tightenlines
,groupaddresses
,prb
,longbibliography
]{revtex4-2}
\pdfoutput=1

\usepackage{graphicx}  
\usepackage{mathtools}
\usepackage{amsmath, amssymb, bm,bbm}   
\usepackage{xcolor}
\usepackage{natbib}
\usepackage{hyperref}

\usepackage{longtable}
\usepackage{csvsimple} 
\usepackage{adjustbox} 
\usepackage{tabularx} 

\usepackage[normalem]{ulem} 

\usepackage{ifthen}

\usepackage{siunitx} 
\usepackage{balance}

\usepackage{orcidlink}

\usepackage{tikz}
\usetikzlibrary{quantikz2}

\newcommand*\xbar[1]{%
   \hbox{%
     \vbox{%
       \hrule height 0.5pt 
       \kern0.3ex
       \hbox{%
         \kern-0.1em
         \ensuremath{#1}%
         \kern-0.1em
       }%
     }%
   }%
} 

\begin{document}

\title{
Training Quantum Dragons}
\author{%
Muhammad Yusf\,\orcidlink{0009-0006-6415-369X}}
\email{mf1478@msstate.edu}

\author{%
Bhola Devkota\, \orcidlink{0009-0001-8633-7805}}
\email{brd189@msstate.edu}

\author{%
Mark A. Novotny\,\orcidlink{0000-0002-1410-5597}}
\email{man40@msstate.edu}

\author{%
Gautam Rupak\, \orcidlink{0000-0001-6683-177X}}
\email{grupak@ccs.msstate.edu}

\affiliation{Department of Physics \& Astronomy and HPC$^2$ Center for 
Computational Sciences, Mississippi State
University, Mississippi State, MS 39762, USA}

\begin{abstract}
The Non-Equilibrium Green's Function (NEGF) is the standard formalism for nano-scale electron transport. By recasting the NEGF scattering problem as a linear system of equations  whose solution encodes the transmission and reflection amplitudes, we present the first quantum computerized implementation of NEGF. We apply both the Harrow--Hassidim--Lloyd  and Variational Quantum Linear Solver algorithms to compute the transmission coefficient $\mathcal{T}(E)$ of quantum dragon nanodevices within the single-band tight-binding model. Quantum dragon devices exhibit perfect transmission across the full conducting band regardless of internal disorder. The problem maps onto compact circuits of 3 and 4 total physical qubits for the 2-site and 6-site dragon devices, respectively. A similarity transformation block-diagonalizes the NEGF linear system reducing  the Pauli decomposition of the block-encoded matrix. We demonstrate the feasibility of quantum computation by performing ideal and noise-aware simulations and computations on physical IBM quantum processor.

\end{abstract}
\maketitle

\section{Introduction}
\label{sec:Intro}


Understanding the electron transport phenomena through nano-scale devices is central to the design of the next-generation electronic components, molecular junctions, and quantum-coherent circuits~\cite{thakur2023, brandbyge2002, DATTA1995}. The standard theoretical framework for such transport calculations is the Non-Equilibrium Green's Function (NEGF) method \cite{thakur2023, DATTA1995}, calculating the electrical conductance from the quantum transmission probability ${\cal T}(E)$ for injected electrons with energy $E$. For non-interacting (free-fermion) systems of size $N$ governed by quadratic tight-binding Hamiltonians, classical NEGF requires solving a single-particle linear system that scales at best as $\mathcal{O}(N^2)$ for structured sparse systems, as $\mathcal{O}(N)$ for one-dimensional chains~\cite{Lewenkopf2013}, and in general as $\mathcal{O}(N^3)$~\cite{harrow2009}. While tractable for small non-interacting devices, this polynomial cost becomes prohibitive at large $N$, and grows exponentially once many-body correlations are introduced~\cite{Golze2019, Georges1996}. Furthermore, even within the non-interacting regime, NEGF on general disordered or large sparse graphs belongs to the Bounded-error Quantum Polynomial time (BQP)-hard complexity class~\cite{Stroeks2024}, and can render the matrix to be inverted highly ill-conditioned. This limits the accuracy and efficiency of classical solvers~\cite{BravoPrieto:2023} which can be an opportunity to test quantum algorithms for a potential advantage.

 Although the NEGF numerical framework for classical computers is well established, its implementation is not directly amenable to quantum computers because it relies on non-unitary matrix inversions and open-boundary complex self-energies that have no natural representation as unitary quantum circuits. To overcome this, we recast the NEGF transport problem as a scattering linear system of equations (LSE) whose solution encodes the transmission and reflection amplitudes of the device~\cite{DATTA1995, NOVO2014, harrow2009}. This reformulation makes the problem directly accessible to quantum linear algebra algorithms.

 Feynman first envisioned a quantum processor that can naturally represent and manipulate quantum states in a Hilbert space whose dimension grows exponentially with the number of qubits, making it a natural platform for simulating quantum mechanical systems~\cite{feynman1982}. This has since crystallized into a broad research program aimed at harnessing quantum computers to solve problems in quantum chemistry, condensed-matter, nuclear and particle physics, and materials science that are intractable on classical computers~\cite{cao2019, mcardle2020}. A landmark algorithmic result in this direction, relevant to this work, is the Harrow--Hassidim--Lloyd (HHL) algorithm~\cite{harrow2009}, which demonstrated that a sparse, well-conditioned $N\times N$ LSE, with a system condition number $\kappa$, can in principle be solved on a quantum computer in $\mathcal{O}[\kappa\,\mathrm{poly}(\log N)]$ time. This is an exponential improvement over the best classical algorithms in the large-$N$ limit. The HHL result established that quantum linear-algebra subroutines could accelerate a wide class of scientific computing tasks, including those arising in electronic-structure theory and transport calculations.

Despite these theoretical promises, practical quantum advantage on near-term hardware remains a significant challenge. Current quantum processors belong to the Noisy Intermediate-Scale Quantum (NISQ) era~\cite{preskill2018}. The NISQ processors operate with $\mathcal{O}(10^2)$--$\mathcal{O}(10^3)$ qubits, but gate fidelities and coherence times are still insufficient for deep fault-tolerant (FT) circuits such as those required by HHL. This has motivated the development of hybrid quantum-classical variational algorithms, which offload some of the computational burden to a classical optimizer while using the quantum processor only for shallow parametrized circuits~\cite{peruzzo2014, mcclean2016, tilly2022, Yusf:2024igb}. The Variational Quantum Eigensolver (VQE)~\cite{peruzzo2014, tilly2022} is the archetype of this approach for eigenvalue problems and has been successfully demonstrated for small molecular systems on several quantum hardware platforms~\cite{google2020}. Analogously, the Variational Quantum Linear Solver (VQLS)~\cite{BravoPrieto:2023} extends the variational philosophy to LSE, minimizing a cost function that vanishes when the trial state coincides with the solution. Importantly, the local cost function variant of VQLS is designed to mitigate barren plateaus-regions of exponentially vanishing gradients that otherwise stall variational optimization at scale~\cite{Cerezo:2021}. These advances in near-term algorithms have renewed interest in applying quantum computers to condensed-matter and materials problems, including lattice models ~\cite{bauer2020, cade2020}.

Here, we present the first quantum computer implementation of NEGF via the HHL and VQLS algorithms to compute the transmission coefficient $\mathcal{T}(E)$ of nano-scale quantum devices modeled within the tight-binding framework. Translationally invariant nanodevices, including graphene nanoribbons and carbon nanotubes, have been measured to exhibit  ballistic transport~\cite{Frank1998, Bolotin2008}, namely $\mathcal{T}(E) = 1$. The $\mathcal{T}(E) = 1$ property has been utilized  to make ballistic FET (Field Effect Transistors) using carbon nanotubes~\cite{javey2003ballistic}, and potentially to utilize graphene to make quality qubits and gates for quantum computers~\cite{Trauzettel2007, Geim2007, Assouline2023}. The devices we choose to consider, known as quantum dragons~\cite{NOVO2014, Inkoom2018, Novotny_2021, Novotny_2023, Novotny2025}, exhibit the remarkable property of perfect transmission $\mathcal{T}(E) = 1$ across the entire attached lead conducting energy band, regardless of internal disorder. Quantum dragon nanodevices may be very disordered---hence no translational invariance---while still exhibiting $\mathcal{T}(E) = 1$.

We examine both a 2-site nonlinear and a 6-site hexagonal quantum dragon, the latter sharing the hexagonal connectivity relevant to graphene and carbon nanotube junctions~\cite{CastroNeto2009, Hamada1992, Nakada1996, Reich2002, Saito1992, NOVO2014, Novotny2025}. We provide the formalism to implement these devices on the qubit registers for quantum computations. While ideal HHL~\cite{harrow2009} simulations achieves $\approx 99\%$ overlap across all energy values, the deep circuits render it highly susceptible to noise on current hardware---a finding that directly motivates the use of VQLS as the near-term algorithm of choice.

We present hybrid quantum-classical VQLS results obtained on an IBM quantum processor (\texttt{ibm\_torino}), as well as noise-aware Qiskit~\cite{Qiskit} simulator results, thereby demonstrating that an optimized implementation of the VQLS algorithm  recovers the perfect-transmission dragon solution and mostly reproduces the wavefunction overlap with the exact classical result. These calculations serve a dual purpose: benchmarking VQLS as a viable near-term algorithm for NEGF-based transport problems, and establishing HHL as the natural FT-era successor, providing a clear pathway toward quantum simulation of more complex nanodevice Hamiltonians as hardware continues to improve.
Crucially, HHL is the natural candidate for FT hardware: its exponential advantage in system size becomes practically accessible once logical error rates are sufficiently suppressed. Recent milestones signal that FT hardware is no longer a distant  prospect: Google's Willow processor has demonstrated exponential suppression of logical error rates below the FT threshold~\cite{googleWillow2024}, and both IBM and Quantinuum project hundreds of logical qubits within this decade~\cite{IBMroadmap2024, bravyi2024highthreshold, quantinuumroadmap2024}. 

The  paper is organized as follows. Section~\ref{sec:device} introduces the nanodevice models and derives the linear systems that encode the transmission coefficient. Sections~\ref{sec:hhl} and~\ref{sec:VQLS} summarize the HHL and VQLS algorithms and their tailored application to the device problem, respectively. In subsection~\ref{sec:trialstate}, we describe the SVD-based ansatz used for trial-wavefunction preparation. Section~\ref{sec:results} presents and analyzes the quantum computational results with both algorithms for the 2-site quantum dragon and with VQLS for the 6-site quantum dragon. Section~\ref{sec:conclusions} offers conclusions, algorithmic limitations, and an outlook for future work.

\section{Nanodevices}
\label{sec:device}

We start with a discussion of particle transport on a wire, before the discussion of nanodevices connected to leads, to introduce the general framework for quantum computation. 
As we show later, the general form of the problem involves solving a LSE
\begin{align}\label{eq:LSE}
A \vec{\bm x}= \vec{\bm b}\,,
\end{align}
 where the matrix  $A$ is defined by the device Hamiltonian and the leads. The vector $\vec{\bm b}$ is determined by the continuity of the scattering wave solution with incident, reflected and transmitted components at the  incident energy $E$. The  components of $\vec{\bm x}$ are used to calculate the reflection $\mathcal{R}(E)$ and transmission $\mathcal{T}(E)$ coefficients as discussed later.

\subsection{Linear 2-site device}
\label{subsec:2siteslinear}

\begin{figure}[htb]
\begin{center}
\includegraphics[width=0.49\textwidth,clip=true]{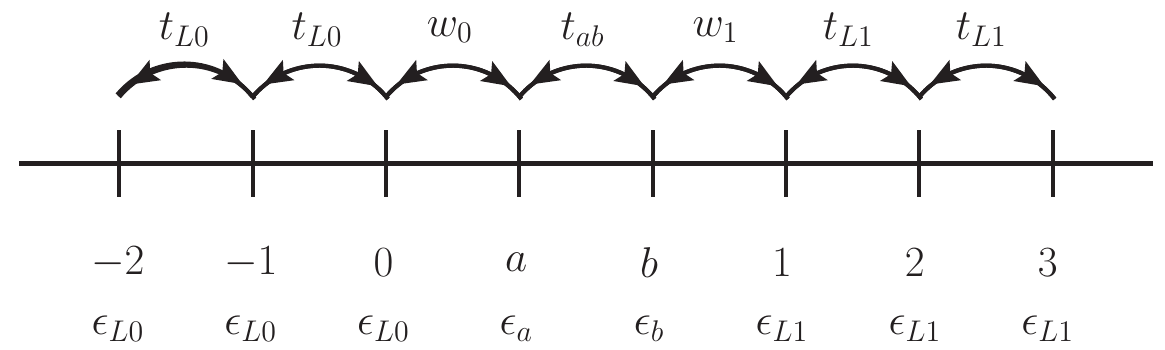}
\end{center}
\caption{\protect  A linear 2-site device. $t_{L0}$, $w_0$, $w_1$, $t_{L1}$ are hopping parameters and $\epsilon_{L0}$, $\epsilon_a$, $\epsilon_b$, $\epsilon_{L1}$ are onsite energies as described in the text.}
\label{fig:2sites}
\end{figure}

Suppose we consider a 2-site device connected to two semi-infinite leads labeled $L0$ and $L1$, respectively. As illustration we use Fig.~\ref{fig:2sites}, where the electrons travel from the left along $L0$ and go out on the right along $L1$. The scattering centers on the incoming lead $L0$ are numbered 0, -1, $\dots$, with onsite potential $\epsilon_{L0}$ and  hopping strength $t_{L0}$ within the sites on $L0$. The scattering centers on $L1$ are labeled 1, 2, $\dots$, with 
 onsite potential $\epsilon_{L1}$ and  hopping strength $t_{L1}$ within the sites on $L1$. The two sites on the device are labeled $a$, $b$ with onsite energies $\epsilon_a$, $\epsilon_b$ and intersite hopping strength $t_{ab}$ as indicated in Fig.~[\ref{fig:2sites}]. This is a ``linear" device. 

We can motivate the hopping strength and the onsite potential as follows. For a particle of mass $m$ in 1-dimension, we write the time-independent Schr\"odinger
\begin{align}
-\frac{\hbar^2}{2m}\frac{d^2}{dx^2}\psi +V(x)\psi= E\psi\,,
\end{align}
which we can approximate on a lattice with spacing $a$ as
\begin{align}
-\frac{\hbar^2}{2m}\frac{\psi(x+a)+\psi(x-a) -2 \psi(x)}{a^2} +V(x)\psi= E\psi\,,
\end{align}
with nearest neighbor finite differences for the derivatives. This describes hopping of the particle from site to site with strength $\hbar^2/(2ma^2)$ and onsite potential $\hbar^2/(ma^2)+V(x)$. Many-body interactions in physical devices would generate an effective mass and onsite potential for electron transport, however, the general form of the Schr\"odinger equation for electron transport in the incoming lead with a   hopping strength $t_{L0}$ and onsite potential $\epsilon_{L0}$ holds. A similar form for the outgoing lead as depicted in Fig.~\ref{fig:2sites} holds. We describe the 2-site device with such a form as well with onsite potentials $\epsilon_a$, $\epsilon_b$, and hopping strength $t_{ab}$ between the two sites.   The incoming lead $L0$ connects to the device at site $a$ with hopping strength $w_0$ and the outgoing lead $L1$ connects to the device at site $b$ with hopping strength $w_1$. 

We seek scattering solutions with incident, reflected and transmitted components such that 
\begin{align}
\psi_n&=e^{iq_0 n} + r e^{-iq_0 n}\,, & n\leq 0\,,\nonumber\\
\psi_n&=se^{i q_1 (n-1)}\,, & n\geq 1\,,
\end{align}
for incoming lead $L0$ and outgoing lead $L1$  with  momenta $q_0$  and $q_1$, respectively,  written in dimensionless units of $\SI{1}{\eV}$ energy. We used the freedom in defining the transmission amplitude $s$ such that $\psi_1=s$.  Then on the incoming $L0$ and outgoing $L1$ leads we get
\begin{align}
&\epsilon_{L1}-E -2t_{L1}\cos q_1=0\,,\nonumber\\
&\left[\epsilon_{L0}-E - 2t_{L0}\cos q_0\right] [e^{iq_0 n}+ r e^{-iq_0 n}]=0\,,
\end{align}
 respectively. This leads to constraints on the incoming $q_0$ and outgoing $q_1$  momenta in relation to the energy $E=\epsilon_{L0}- 2t_{L0}\cos q_0=\epsilon_{L1}-2t_{L1}\cos q_1$. Thus propagating waves have to satisfy
\begin{align}\label{eq:propagating}
\epsilon_{L0}-2t_{L0}\leq E\leq\epsilon_{L0}+2t_{L0}\,, \nonumber\\
\epsilon_{L1}-2t_{L1}\leq E\leq\epsilon_{L1}+2t_{L1}\,.
\end{align}
 
 We write the wavefucntion at sites $a$ and $b$  as $\psi_a$ and $\psi_b$, respectively. The interaction of the leads $L0$, $L1$ with the 2-site device then requires interactions between sites 0, $a$, $b$, and 1 described by 4 linear equations 
 \begin{align}\label{eq:2site}
-t_{L0}\psi_{-1} +(\epsilon_{L0}-E)\psi_0-w_0\psi_a =0\,,\nonumber\\
-w_0\psi_0+(\epsilon_a-E)\psi_a-t_{ab}\psi_b=0\,,\nonumber\\
-t_{ab}\psi_a+(\epsilon_b-E)\psi_b-w_1\psi_1=0\,,\nonumber\\
-w_1\psi_b+(\epsilon_{L1}-E)\psi_1 -t_{L1}\psi_2=0\,,
\end{align}
in terms of wavefunctions at 6 sites to be solved. However, using the scattering solutions $\psi_n$ for $n\leq 0$ and $n \geq 1$ we can write the linear system of 4 equations
\begin{multline}\label{eq:2siteAxb}
\begin{pmatrix}
\xi_0-E &-w_0 & 0 &0\\
-w_0&\epsilon_a-E & -t_{ab} &0\\
0 &-t_{ab} &\epsilon_{b}-E &-w_1\\
0&0&-w_1 &\xi_1-E
\end{pmatrix}\begin{pmatrix}
1+r\\ \psi_a \\ \psi_b \\s
\end{pmatrix}\\=\begin{pmatrix}
-i 2 t_{L0}\sin q_0\\ 0 \\0 \\0
\end{pmatrix}\,,
\end{multline}
in terms of the reflection $r$ and transmission $s$ coefficients, reducing the number of unknowns to just 4: $r$, $\psi_a$, $\psi_b$, $s$ with 
\begin{align}\label{eq:leadenergy}
\xi_j &= \frac{\epsilon_{Lj}+E}{2}-i \frac{\sqrt{4 t_{Lj}^2-(\epsilon_{Lj}-E)^2}}{2}\nonumber\\
&=\frac{\epsilon_{Lj}+E}{2}-i t_{Lj}\sin q_j\,,
\end{align}
for $j=0,1$. 
We solve Eq.~(\ref{eq:2siteAxb}) for the transmission coefficient
\begin{align}
&\mathcal T(E)=\frac{v_0}{v_1}|s|^2\,, & &\nonumber\\
&v_0=\frac{dE}{d q_0}=2 t_{L0}\sin{q_0}\, , & v_1&=\frac{dE}{dq_1}=2 t_{L1}\sin{q_1}\,.
\end{align}
For the device problems, we choose the two leads to be identical---$t_{L0}=t_{L1}$ and $\epsilon_{L0}=\epsilon_{L1}$---such that the group velocities $v_0=v_1$ and momenta $q_0=q_1$ are the same, and $\mathcal T(E)=|s|^2$.

\subsection{2-site Quantum Dragons}
\label{subsec:2siteDragon}

\begin{figure}[htb]
\begin{center}
\includegraphics[width=0.49\textwidth,clip=true]{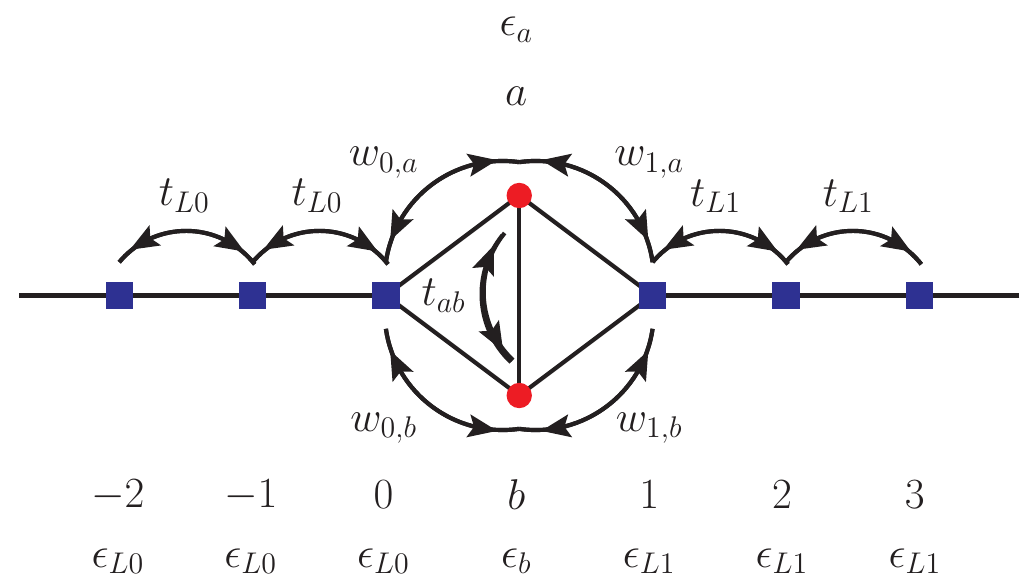}
\end{center}
\caption{\protect  2-site quantum dragon. Hopping parameters and onsite energies as indicated.}
\label{fig:2sitesDragon}
\end{figure} 
The first device we calculate using quantum computing algorithms is the 2-site ``quantum dragon"~\cite{NOVO2014}. This can be viewed as a ``nonlinear" device with a more complete hopping between the two leads $L0$ and $L1$ to the two sites as shown in Fig.~\ref{fig:2sitesDragon}. The LSE to solve for transmission for such a nonlinear device is a more generalized form of Eq~(\ref{eq:2siteAxb}):
\begin{multline}\label{eq:2siteNonLinearDevice}
\begin{pmatrix}
\xi_0-E &-w_{0,a} & -w_{0,b} &0\\
-w_{0,a} & \epsilon_a-E & -t_{ab} & -w_{1,a} \\
-w_{0,b}  & -t_{ab} &\epsilon_{b}-E & -w_{1,b} \\
0 & -w_{1,a} & -w_{1,b} &\xi_1-E
\end{pmatrix}\begin{pmatrix}
1+r\\ \psi_a \\ \psi_b \\s
\end{pmatrix}\\=\begin{pmatrix}
-i 2 t_{L0}\sin q_0\\ 0 \\0 \\0
\end{pmatrix}\,,
\end{multline}
where there can be electron transport from both the leads to either of the two sites on the device.
$\xi_j$  is still given by Eq.~(\ref{eq:leadenergy}). 
The hopping onto the two sites of the device from the incoming lead  $L0$ have strength $w_{0,a}$ 
and $w_{0,b}$.  
The hopping from the two sites of the device to the outgoing lead $L1$ have strength $w_{1,a}$ 
and $w_{1,b}$.  We again choose the leads to be composed of the same material and thus the 
hopping parameters $t_{L0}=t_{L1}\equiv t_L$, $w_{0,a}=w_{1,a}\equiv w_a$, $w_{0,b}=w_{1,b}\equiv w_b$, onsite energies $\epsilon_{L0}=\epsilon_{L1}\equiv \epsilon_L$, momenta $q_0=q_1\equiv q$, and energies $\xi_0=\xi_1\equiv\xi$ are the same. The values of the parameters can be chosen at random to test our quantum computing algorithms. Some explicit values are given in Appendix~\ref{app:dragon_params}.

The quantum dragon solution~\cite{NOVO2014} for this device where the transmission coefficient $\mathcal T(E)=1$ for 
$\epsilon_L-2 t_L\leq E\leq \epsilon_L+2 t_L$ is obtained with the following choice of device parameters:
\begin{align}\label{eq:vdragon}
\begin{pmatrix}
w_{a}\\w_{b}\end{pmatrix}
=t_{L}\begin{pmatrix} \cos\beta\\\sin\beta\end{pmatrix}\equiv \vec{v}_\text{dragon}\,,
\end{align}
with $0<\beta<\pi/2$ for device-lead hopping and device Hamiltonian
\begin{align}\label{eq:dragon_cond_A}
\begin{pmatrix}
\epsilon_{a} & -t_{ab}\\ -t_{ab} &\epsilon_b
\end{pmatrix} \vec{v}_\text{dragon} = \epsilon_L \vec{v}_\text{dragon}\,.
\end{align}
Explicitly, this requirement is 
\begin{align}
\epsilon_{a}&= t_{ab}\tan\beta +\epsilon_L\,, & \epsilon_{b}&= t_{ab}\cot\beta +\epsilon_L\,.
\end{align}
Thus, for fixed lead parameters $\epsilon_L$ and $t_L$, any arbitrary $0<\beta<\pi/2$ and $t_{ab}$ supports the quantum dragon solution with perfect transmission  $\mathcal T(E)=1$ in the allowed energy region for scattering. 

A convenient way to rewrite the linear system of equations for the dragon solution is 
\begin{multline} \label{eq:2siteDragon}
\begin{pmatrix}
\xi-E & -\vec{v}_\text{dragon}^\dagger & 0\\
-\vec{v}_\text{dragon} & {\bf A} +(\epsilon_L-E){\bf I} &-\vec{v}_\text{dragon}\\
0 &-\vec{v}_\text{dragon}^\dagger &\xi-E
\end{pmatrix}
\begin{pmatrix}
1+r\\ \vec{\bm\Psi}_1\\ s
\end{pmatrix}\\
=\begin{pmatrix}
-i2 t_L\sin q\\
\vec{\bm{0}}\\
0
\end{pmatrix}\,,
\end{multline}
where we defined
\begin{align}
{\bf A}&\!=\! \begin{pmatrix}
\epsilon_{a} -\epsilon_L& -t_{ab}\\ -t_{ab} &\epsilon_b-\epsilon_L
\end{pmatrix}\!=\! t_{ab}\!\begin{pmatrix} \tan\beta &-1\\ -1& \cot\beta \end{pmatrix}\,, \nonumber\\
 \vec{\bm\Psi}_1&=\begin{pmatrix}\psi_a \\\psi_b\end{pmatrix}\,,
\end{align}
and ${\bf I}$ is a $2\times2$ identity matrix and $\vec{\bf 0}$ is a 2-dimensional null vector. 

The  quantum dragon  condition~\cite{NOVO2014,Inkoom2018,Novotny_2021,Novotny_2023} for $\mathcal T(E)=1$ solution can be stated as an eigenvalue equation 
\begin{align}\label{eq:dragon_cond_B}
{\bf A} \vec{v}_\text{dragon}= 0\,.
\end{align}
To make the perfect transmission solution apparent we perform similarity transformations~\cite{NOVO2014, Novotny2025} with
\begin{align}\label{eq:2site_similarity}
\mathcal U_4&=\begin{pmatrix}1 &\vec{\bm 0}^\dagger &0\\
\vec{\bm 0} & {\bf U} &\vec{\bm 0} \\
0 & \vec{\bm 0}^\dagger & 1
\end{pmatrix}\,, & {\bf U} &= \begin{pmatrix} \cos\beta &\sin\beta \\ \sin\beta &-\cos\beta \end{pmatrix}\,,
\end{align}
where ${\bf U} = {\bf U}^\dagger={\bf U}^{-1}$ is both Hermitian and unitary, and consequently $\mathcal U_4=\mathcal U_4^\dagger=\mathcal U_4^{-1}$ as well. Under a similarity transformation of ${\bf M}_4$ which is the $4\times 4$ matrix appearing on the left side of Eq.~(\ref{eq:2siteDragon}), we find:
\begin{align} \label{eq:2siteDragon:02}
\mathcal U_4^{-1}{\bf M}_4\mathcal U_4=\begin{pmatrix}
\xi-E &-t_L &0 &0
\\-t_L &\epsilon_L-E&0 &0
\\0 &0&\epsilon_L-E+\eta&0
\\ 0 &-t_L &0 &\xi -E
\end{pmatrix}\,,
\end{align}
where we defined $\eta= t_{ab}\sec\beta\csc\beta$ and used the relations
\begin{align}
{\bf U}^{-1} {\bf A}{\bf U}&=\eta \!\begin{pmatrix} 0 & 0\\ 0 &1 \end{pmatrix}\,, &
{\bf U}^{-1} \vec{v}_\text{dragon}&= t_L \!\begin{pmatrix} 1\\0 \end{pmatrix}\,.
\end{align}
$\mathcal U_4^{-1}{\bf M}_4\mathcal U_4$ can be cast in a block diagonal form by interchanging the 3rd-4th row-column with the permutation matrix
\begin{align}
{\bf P}_4=\begin{pmatrix} 1 & 0 & 0 & 0\\
0 & 1& 0 &0\\
0& 0 & 0&1\\
0& 0& 1 &0
\end{pmatrix}\,,
\end{align}
for the permutation cycle $(3\,4)$ such that
\begin{multline}
{\bf P}_4^{-1} \mathcal U_4^{-1}{\bf M}_4\mathcal U_4{\bf P}_4\\
=\begin{pmatrix}
\xi-E & -t_L &0 &0
\\ -t_L &\epsilon_L-E & -t_L &0
\\0 & -t_L &\xi-E &0
\\ 0 & 0 & 0 & \epsilon_L -E +\eta
\end{pmatrix}\,.
\end{multline}
Thus the LSE can be written as 
\begin{widetext}
\begin{align}\label{eq:2siteDragonAxb}
&{\bf M}_4\begin{pmatrix}
1+r\\ \vec{\bm\Psi}_1\\ s
\end{pmatrix}
=\begin{pmatrix}
-i2 t_L\sin q\\
\vec{\bm{0}}\\
0
\end{pmatrix}  
\Rightarrow 
{\bf P}_4^{-1} \mathcal U_4^{-1}{\bf M}_4\mathcal U_4{\bf P}_4{\bf P}_4^{-1}\mathcal U_4^{-1}
\begin{pmatrix}
1+r\\ \vec{\bm\Psi}_1\\ s
\end{pmatrix} 
={\bf P}_4^{-1}\mathcal U_4^{-1}
\begin{pmatrix}
-i2 t_L\sin q\\ \vec{\bf 0} \\0
\end{pmatrix} &\nonumber\\
\Rightarrow &
\begin{pmatrix}
\xi-E & -t_L &0 &0
\\ -t_L &\epsilon_L-E & -t_L &0
\\0 & -t_L &\xi-E &0
\\ 0 & 0 & 0 & \epsilon_L+\eta -E 
\end{pmatrix}
\begin{pmatrix} 1+r 
\\ \psi_{a}\cos\beta+\psi_{b}\sin\beta 
\\s
\\ \psi_{a}\sin\beta-\psi_{b}\cos\beta
\end{pmatrix}
=\begin{pmatrix}
-i2 t_L\sin q\\ \vec{\bf 0} \\0
\end{pmatrix} 
\,.
\end{align}
Comparing the above with Eq.~(\ref{eq:2siteAxb}),
we see that the wavefunction $\psi_{+}\equiv\psi_{a}\cos\beta+\psi_{b}\sin\beta$ describes  electron transport in  a single site linear device that is decoupled from the  $\psi_{-}\equiv\psi_{a}\sin\beta-\psi_{b}\cos\beta$ wavefunction which satisfies  $\psi_{-}=0$. 
Moreover, since the hopping between the leads and the device with wavefunction $\psi_+$ have by the same hopping parameter $t_L$ it results in a perfect transmission. 
The explicit values for the parameters 
students for the 2-site quantum dragons 
are given in Appendix~\ref{app:dragon_params}.  
\end{widetext}

\subsection{6-site Quantum Dragons}
\label{subsec:6siteDragon}

The second nanodevice we consider is a 6-site quantum dragon. The particular connections we choose for the device sites in the implementation for quantum circuits can be described as a hexagon as we will explain below.  Both graphene and carbon nanotubes 
have an underlying hexagonal graph, with a hopping 
strength of about $2.7$~eV 
\cite{CastroNeto2009,Reich2002,Saito1992}.  
Connecting armchair single-walled carbon 
nanotubes (SWCNTs) to our thin leads makes the 
hopping strength in the leads the same as 
the hopping strength of the carbon-carbon bond 
\cite{NOVO2014}.

Tight binding Hamiltonians have also been used to study 
electron transport in DNA 
\cite{porath2000direct,2005Gutierrez,2007Cuniberti,2023Hashem},
with the strength of the hopping depending on the particular base pairs, 
but in the range of 0.1~eV to 0.4~eV.  

\begin{figure}[htb]
\begin{center}
\includegraphics[width=0.49\textwidth,clip=true]{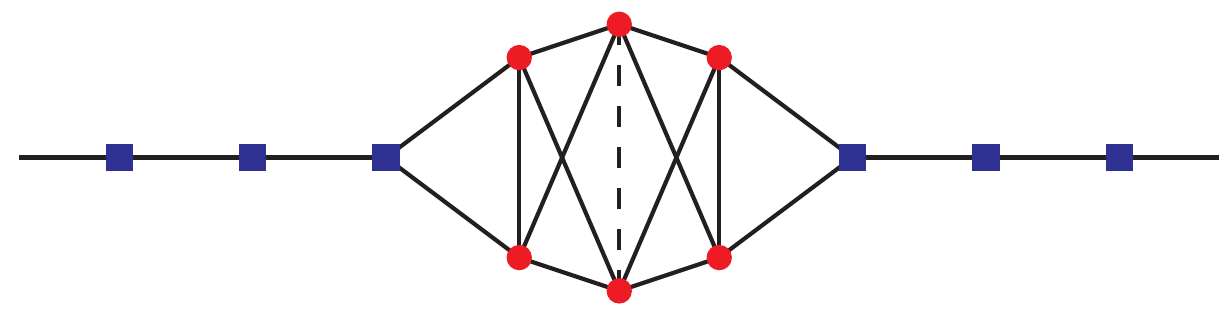}
\end{center}
\caption{\protect  6-site quantum dragon. The 6 sites on the device are arranged in 3 slices left-to-right with two sites per slice.}
\label{fig:6sitesDragon}
\end{figure} 

For the 6-site device, we generalize the construction from the 2-site quantum dragon as shown in Fig.~\ref{fig:6sitesDragon}, where we arranged the sites into 3 slices labeled $l=1$, 2 and $3$ with $m=2$ sites per slice. We remove the hopping within the two sites for the second slice $l=2$ later. The 6 sites on the nanodevice shown as red dots form a hexagon. 
Generalizing the previous notation, we derive the 6-site LSE as
\begin{widetext}
\begin{multline} \label{Eq:6siteDragon:01}
\begin{pmatrix}
\xi-E & -\vec{v}_\text{dragon,\,1}^\dagger &\vec{\bm 0}^\dagger &\vec{\bm 0}^\dagger& 0\\ 
-\vec{v}_\text{dragon,\,1} & {\bf A}_1 +(\epsilon_L-E){\bf I} & {\bf B}_{1,2} &{\bf 0} &\vec{\bm 0}\\ 
\vec{\bf 0} &{\bf B}_{12}^\dagger &  {\bf A}_2+(\epsilon_L-E){\bf I} & {\bf B}_{2,3} & \vec{\bm 0}\\ 
\vec{\bf 0} & {\bf 0} &{\bf B}_{23}^\dagger &  {\bf A}_2+(\epsilon_L-E){\bf I} & -\vec{v}_\text{dragon,\,3}\\ 
0 &\vec{\bm 0}^\dagger &\vec{\bm 0}^\dagger &-\vec{v}_\text{dragon,\,3}^\dagger &\xi-E 
\end{pmatrix}
\begin{pmatrix}
1+r\\ \vec{\bm\Psi}_1\\ \vec{\bm\Psi}_2\\ \vec{\bm\Psi}_3\\ s
\end{pmatrix}
=\begin{pmatrix}
-i2 t_L\sin q\\
\vec{\bm{0}}\\ \vec{\bm{0}}\\ \vec{\bm{0}}\\
0
\end{pmatrix}\,,
\end{multline}
\end{widetext}
where ${\bf 0}$ is a $2\times2$ zero matrix which is distinct from the 2-dimensional null vector $\vec{\bf 0}$. 
The dragon vector defined as
\begin{align}
\vec{v}_\text{dragon,\,j}= t_L\begin{pmatrix} \cos\beta_j\\ \sin\beta_j \end{pmatrix}\,,
\end{align}
satisfies the quantum dragon  eigenvalue equation  ${\bf A}_j \vec{v}_\text{dragon,\,j}=\epsilon_L\vec{v}_\text{dragon,\,j}$ for hopping within each slice for
\begin{align}
{\bf A}_j &= t_\text{intra,j}\begin{pmatrix} \tan\beta_j &-1\\-1 &\cot\beta_j \end{pmatrix}\,.
\end{align}
with the 2-site wavefunction at each slice $j$ grouped together as
\begin{align}
\vec{\bm \Psi}_j=\begin{pmatrix} \psi_{1,j} \\ \psi_{2,j}\end{pmatrix}\,.
\end{align}
$t_\text{intra,j}$ is the hopping parameter between the two sites within slice $j$. This is identical to the derivation in Eqs.~(\ref{eq:dragon_cond_A}) and (\ref{eq:dragon_cond_B}). 
We specify hopping between neighboring slices $j$ and $j+1$ by the matrices
\begin{align}
{\bf B}_{j,j+1}=\begin{pmatrix}
t_{(j,a),(j+1,a)} & t_{(j,a),(j+1,b)} \\
t_{(j,b),(j+1,a)} & t_{(j,b),(j+1,b)} 
\end{pmatrix}\,,
\end{align}
where we labeled the two sites in a slice $a$ and $b$ similar to the 2-site quantum dragon. $t_{(j,m),(j+1,n)} $ is the hopping between site $m$ of slice $j$ and site $n$ of slice $j+1$. The quantum dragon condition for inter-slice hopping ${\bf B}_{j,j+1}\vec{v}_\text{dragon,\,j}=-t_L\vec{v}_\text{dragon,\,j}$~\cite{NOVO2014,Inkoom2018,Novotny_2021} is obtained for the parameterization
\begin{widetext}
\begin{align}\label{eq:Bjj1}
B_{j,j+1}=\begin{pmatrix} \cos\beta_j &\sin\beta_j\\ \sin\beta_j&-\cos\beta_j\end{pmatrix}
\begin{pmatrix}
-t_L &0\\0 &-\kappa_j
\end{pmatrix}
\begin{pmatrix} \cos\beta_{j+1} &\sin\beta_{j+1}\\ \sin\beta_{j+1}&-\cos\beta_{j+1}\end{pmatrix}
\,,
\end{align}
where the singular value $|\kappa_j|\leq1$ is obtained when
\begin{align}
\label{Eq:6site:General:05}
-{\rm min}\left(\tan\left(\beta_{j+1}\right)\tan\left(\beta_{j}\right),
\cot\left(\beta_{j+1}\right)\cot\left(\beta_{j}\right)\right) 
\leq \kappa_j \leq
{\rm min}\left(\tan\left(\beta_{j+1}\right)\cot\left(\beta_{j}\right),
\cot\left(\beta_{j+1}\right)\tan\left(\beta_{j}\right)\right)\,.
\end{align}
\end{widetext}

Similar to the 2-site quantum dragon device Eq.~(\ref{eq:2site_similarity}), we construct Hermitian and unitary matrix
\begin{align}\label{eq:6site_similarity}
\mathcal U_8&=\begin{pmatrix}1 &\vec{\bm 0}^\dagger &\vec{\bm 0}^\dagger &\vec{\bm 0}^\dagger &0\\ 
\vec{\bm 0} & {\bf U}_1 &{\bm 0}  &{\bm 0} &\vec{\bm 0} \\ 
\vec{\bm 0} &{\bm 0}  & {\bf U}_2  &{\bm 0} &\vec{\bm 0} \\ 
\vec{\bm 0} &{\bm 0}  &{\bm 0} & {\bf U}_3 &\vec{\bm 0} \\ 
0 & \vec{\bm 0}^\dagger  & \vec{\bm 0}^\dagger  & \vec{\bm 0}^\dagger  & 1 
\end{pmatrix}\,, \nonumber\\
 {\bf U}_j &= \begin{pmatrix} \cos\beta_j &\sin\beta_j \\ \sin\beta_j &-\cos\beta_j \end{pmatrix}\,,
\end{align}
for a similarity transformation of the $8\times8$ matrix ${\bf M}_8$ that appear on the left hand side of Eq.~(\ref{Eq:6siteDragon:01}). We also construct a permutation operator ${\bf P}_8^{-1}$ to shuffle the wavefunctions $\psi_{j,+}=\psi_{1,j}\cos\beta_j+\psi_{2,j}\sin\beta_j$ above  and the 
$\psi_{j,-}=\psi_{1,j}\sin\beta_j-\psi_{2,j}\cos\beta_j$ 
 wavefunctions to the bottom such that
\begin{align}
{\bf P}_8^{-1} \mathcal U_8 \begin{pmatrix} 1+r \\ \bm{\Psi}_1 \\ \bm{\Psi}_2 \\ \bm{\Psi}_3 \\ s\end{pmatrix}=
{\bf P}_8^{-1}   \begin{pmatrix} 1+r  \\ \psi_{1,+} \\ \psi_{1,-} \\ \psi_{2,+}\\ \psi_{2,-} \\ \psi_{3,+} \\ \psi_{3,-}  \\s\end{pmatrix}=
\begin{pmatrix} 1+r  \\ \psi_{1,+} \\ \psi_{2,+} \\ \psi_{3,+}\\ s \\ \psi_{1,-} \\ \psi_{2,-} \\ \psi_{3,-} \end{pmatrix}\,.
\end{align}
This is accomplished with the inverse-permutation operator 
\begin{align}
\label{Eq:MAN:6site:41}
{\bf P}_8^{-1} =
\begin{pmatrix}
1 & 0 & 0 & 0 & 0 & 0 & 0 & 0 \\
0 & 1 & 0 & 0 & 0 & 0 & 0 & 0 \\
0 & 0 & 0 & 1 & 0 & 0 & 0 & 0 \\
0 & 0 & 0 & 0 & 0 & 1 & 0 & 0 \\
0 & 0 & 0 & 0 & 0 & 0 & 0 & 1 \\
0 & 0 & 1 & 0 & 0 & 0 & 0 & 0 \\
0 & 0 & 0 & 0 & 1 & 0 & 0 & 0 \\
0 & 0 & 0 & 0 & 0 & 0 & 1 & 0 \\
\end{pmatrix}\,,
\end{align}
for permutation cycles  $(3\,6\,4)(5\,7\,8)$.  Notice that
\begin{align}
\begin{pmatrix} \psi_{j,+} \\ \psi_{j,-} \end{pmatrix}
={\bf U}_j \begin{pmatrix} \psi_{1,j}\\ \psi_{2,j}\end{pmatrix}\,.
\end{align}

Performing similarity transformation on ${\bf M}_8$ followed by permutation we get
\begin{widetext}
\begin{align}\label{eq:6siteDragonAxb}
&{\bf M}_8\begin{pmatrix}
1+r\\ \vec{\bm\Psi}_1 \\ \vec{\bm\Psi}_2 \\ \vec{\bm\Psi}_3\\ s
\end{pmatrix}
=\begin{pmatrix}
-i2 t_L\sin q\\
\vec{\bm{0}}\\
0
\end{pmatrix}  
\Rightarrow 
{\bf P}_8^{-1} \mathcal U_8^{-1}{\bf M}_8\mathcal U_8{\bf P}_8{\bf P}_8^{-1}\mathcal U_8^{-1}
\begin{pmatrix}
1+r\\ \vec{\bm\Psi}_1 \\ \vec{\bm\Psi}_2 \\ \vec{\bm\Psi}_3\\ s
\end{pmatrix} 
={\bf P}_8^{-1}\mathcal U_8^{-1}
\begin{pmatrix}
-i2 t_L\sin q\\ \vec{\bf 0} \\ \vec{\bf 0}  \\ \vec{\bf 0} \\0
\end{pmatrix} &\nonumber\\
\Rightarrow &
\begin{pmatrix}
\xi-E & -t_L &0 &0 &0 &0 &0 &0\\ 
 -t_L &\epsilon_L-E & -t_L &0 &0 &0 &0 &0\\ 
0 & -t_L &\epsilon_L-E &-t_L &0  &0 &0 &0\\ 
0 &0 & -t_L &\epsilon_L-E &-t_L &0  &0 &0\\ 
0 &0 &0 & -t_L &\xi-E &0  &0 &0\\ 
0 & 0 & 0 &0 &0 & \epsilon_L +\eta_1-E & -\kappa_1 &0\\ 
0 & 0 & 0 &0 &0 &-\kappa_1 & \epsilon_L +\eta_2-E & -\kappa_2\\ 
0 & 0 & 0 &0 &0 &0 &-\kappa_2 & \epsilon_L +\eta_3-E  
\end{pmatrix}\!
\begin{pmatrix} 1+r  \\ \psi_{1,+} \\ \psi_{2,+} \\ \psi_{3,+}\\ s \\ \psi_{1,-} \\ \psi_{2,-} \\ \psi_{3,-} \end{pmatrix}\
\!=\!\begin{pmatrix}
-i2 t_L\sin q\\0 \\0  \\0 \\0 \\0 \\0 \\0
\end{pmatrix} 
\,.
\end{align}
Again, comparing the above with Eq.~(\ref{eq:2siteAxb}), one concludes that 
the $\psi_{j,+}$  wavefunctions describe  electron transport in  a 3-site linear device that is decoupled from the 
 $\psi_{j,-}$ wavefunctions.  Since the hopping between the leads and the device  sites with wavefunction $\psi_{j,+}$  have  same hopping parameter $t_L$ it results in perfect transmission. The decoupled $\psi_{j,-}$ satisfies $\psi_{j,-}=0$.
 The explicit values of the parameters 
 studied for the 6-site quantum dragon 
 are given in Appendix~\ref{app:dragon_params}.  
\end{widetext}

\subsection{Quantum state}
\label{subsec:QuantumLSE}

In Eqs.~(\ref{eq:2siteDragonAxb}) and  (\ref{eq:6siteDragonAxb}), the calculation of the transmission coefficient $\mathcal{T}(E)=|s|^2$ is reduced to solving a linear system of equation as promised earlier.  To evaluate $s$ on a quantum computer,  the vectors $\vec{\bm x}$ and $\vec{\bm b}$ have to reside on the quantum registers of qubits as normalized state vectors. Thus, we work with the following rescaled vectors  and matrices instead 
\begin{align}\label{eq:QLSE}
 |x\rangle\equiv&\begin{pmatrix}
     x_r \\\vec{\bm x}_+\\x_s\\\vec{\bm x}_-
 \end{pmatrix}=\frac{1}{c} \begin{pmatrix} 1+r\\|\bm\Psi_{+}\rangle \\s \\|\bm\Psi_{-}\rangle \end{pmatrix}\,, & |b\rangle =&\begin{pmatrix}1 \\0 \\ \vdots \\0 \end{pmatrix}\,,\nonumber\\
 \xbar{A}=& \frac{i c }{2 t_L \sin q}A\,, & &
\end{align}
such that $\xbar{A}|x\rangle=|b\rangle$ with $c>0$ which holds for the dragon solution. $|\bm\Psi_{\pm}\rangle$ is a vector comprised of the states $\psi_{j,\pm}$ for various slices $j=1,2,\cdots$.
The normalization constant $c$ is used in post processing to relate the normalized quantum state $|x\rangle$ to the component $s=c x_s$ needed for $\mathcal{T}(E)=|s|^2$. The normalization $c$ can be determined simply once current conservation 
\begin{align}\label{eq:constraint}
|r|^2+|s|^2=1\,,
\end{align}
is imposed to write
\begin{align}
c=\frac{x_r+x_r^\ast}{|x_r|^2+|x_s|^2}\,.
\end{align}

One should note that  both $A$ and the rescaled matrix $\xbar{A}$ are non-Hermitian. This affects how the quantum circuits are constructed in the two algorithms---HHL and VQLS---that we discuss next.

\section{HHL}
\label{sec:hhl}

HHL~\cite{harrow2009} is the first quantum algorithm that we applied to calculate $\mathcal{T}(E)$. It is a 
 quantum phase estimation (QPE) based algorithm for solving linear system of equations $A\vec{\bm x}=\vec{\bm b}$. HHL provides an exponential improvement in system size under the assumptions  that $A$ is sparse and efficiently row-queryable, that $|b\rangle$ can be prepared in $\mathcal{O}[\mathrm{poly}(\log N)]$ time, and that the condition number $\kappa$ is small---the runtime scales as $\mathcal{O}[\kappa\,\mathrm{poly}(\log N)]$, so ill-conditioned systems erode the advantage. The algorithm produces a quantum state $|x\rangle$ proportional to the solution vector $\vec{\bm x}$ which can be read out without incurring an $\mathcal{O}(N)$ measurement overhead that would otherwise negate the quantum speedup~\cite{harrow2009, Aaronson2015}.

 HHL requires that the Hamiltonian matrix be Hermitian, which is not true for $\xbar{A}$. However, this is easily remedied doubling the size of the Hilbert space and embedding the problem in this higher-dimensional space by block-encoding as follows
 \begin{align}\label{eq:Atilde}
     |\tilde{x}\rangle=&\begin{pmatrix}
         0\\|x\rangle
     \end{pmatrix}\,, & |\tilde{b}\rangle=&\begin{pmatrix}
         |b\rangle\\ 0
     \end{pmatrix}\,,\nonumber\\
     \tilde{A}=&\begin{pmatrix}
         0 & \xbar{A}\\ \xbar{A}^\dagger &0
     \end{pmatrix}\,, & &
 \end{align}
such that $\tilde{A}^\dagger=\tilde{A}$, $\tilde{A}|\tilde x\rangle=|\tilde b\rangle$ and correspondingly $\xbar{A}|x\rangle=|b\rangle$ as well. 

Next, we briefly outline the HHL algorithm skipping several details since it and QPE are fairly standard algorithms in quantum computation. This discussion is based on Ref.~\cite{Zaman2023}. The key observations in the HHL algorithm are as follows. The Hermitian matrix $\tilde{A}$ has a spectral decomposition 
\begin{align}
\tilde{A}=\sum_{i=0}^{N_b-1}\lambda_i|\lambda_i\rangle\langle\lambda_i|\,,
\end{align}
where $\tilde{A}|\lambda_i\rangle=\lambda_i|\lambda_i\rangle$ with real eigenvalue $\lambda_i$, and $N_b$ is the dimension of $\tilde{A}$. The solution vector is simply
\begin{align}
|\tilde x\rangle=\tilde{A}^{-1}|\tilde b\rangle= \sum_{i=0}^{N_b-1}\frac{1}{\lambda_i}|\lambda_i\rangle\langle\lambda_i|\tilde b\rangle\equiv
\sum_{i=0}^{N_b-1}\frac{\tilde{b}_i}{\lambda_i}|\lambda_i\rangle\,,
\end{align}
where $\tilde{b}_j=\langle\lambda_j|b\rangle$.
HHL achieves this through a three step process. 

Starting from an initial state $|\tilde b\rangle$ stored in the quantum register of the $b$-block of $n_b=\log_2(N_b)$ qubits, QPE is applied for the spectral decomposition. It involves controlled application of  $U(t)=\exp(i \tilde{A} t)$ to the state $|\tilde{b}\rangle$ with a suitable parameter $t$ we discuss below. The multiple applications of $U(t)$ is controlled by a second set of $c$-block of $n_c=\log_2(N_c)$ counting  qubit registers $|c\rangle$. Given an eigenstate $|\lambda_i\rangle$ such that  $U(t)|\lambda_i\rangle=\exp(i\lambda_i t)|\lambda_i\rangle$, QPE uses phase kick-back to produce the state $|\lambda_i\rangle\otimes|N_c t \lambda_i/(2\pi)\rangle$ to the nearest integer representation of $0\le N_c t \lambda_i/(2\pi)\le N_c-1$. Thus QPE produces the state
\begin{align}
\sum_{i=0}^{N_b-1} \tilde{b}_i|\lambda_i\rangle_b\otimes |N_c t\frac{ \lambda_i}{2\pi}\rangle_c\,,
\end{align}
where the subscript on the kets indicate the respective $b$- and $c$-block quantum registers. 

The second step is a Grover rotation on a single ancilla flag-qubit $|0\rangle_a$ conditioned on the $c$-block register through controlled rotation such that  
\begin{multline}\label{eq:grover}
\sum_{i=0}^{N_b-1} \tilde{b}_i|\lambda_i\rangle_b\otimes |N_c t\frac{ \lambda_i}{2\pi}\rangle_c\otimes|0\rangle_a
\rightarrow
\sum_{i=0}^{N_b-1} \tilde{b}_i|\lambda_i\rangle_b\\\otimes |N_c t\frac{ \lambda_i}{2\pi}\rangle_c\otimes
 \left[\sqrt{1-\frac{C^2}{\lambda_i^2}}|0\rangle_a+\frac{C}{\lambda_i}|1\rangle_a \right]\,,
\end{multline}
where the constant $|C|\leq|\lambda_i|$ depends on the exact rotation one performs. A priori the optimal value of $C$ to increase the probability of the $|1\rangle_a$ state might not be known,  without knowing the spectrum $\lambda_i$. In Appendix~\ref{appendix:tuning_t} we discuss optimal choice of $t$. In this work we employ the builtin \texttt{ExactReciprocalGate()} available in Qiskit~\cite{Qiskit}, which performs the eigenvalue inversion 
\begin{align}\label{eq:exactreciprocal}
   |\theta \rangle_c\!\otimes \!|0\rangle_a\rightarrow|\theta\rangle_c\!\otimes \!
\left[\sqrt{1\!-\!s^2\frac{N_c}{\theta^2}}|0\rangle_a\!+\!s\frac{N_c}{\theta}|1\rangle_a \right]\,,
\end{align}
where $s$ is a scaling factor. 
This makes the inversion exact for any eigenvalue representable within $n_c$ clock qubits,  at the cost of a deeper ancilla rotation sub-circuit. 
 While this removes the spectral commensurate assumption and broadens applicability to arbitrary Hermitian matrices---including the complex dragon matrix arising from the open quantum transport problem studied here---the increased circuit depth amplifies the susceptibility to decoherence and gate errors on near-term hardware, further motivating the use of VQLS for noisy device simulations.

After the eigenvalue inversion, the HHL algorithm  requires a measurement of the flag-qubit, and proceeds only if it results in the  $|1\rangle_a$ state. This results in a state with the desired eigenvalue reciprocal amplitude
\begin{align}
\left[ \sum_{l=0}^{N_b-1} \frac{|\tilde{b}_l|^2}{|\lambda_l|^2}\right]^{-\frac{1}{2}}
\sum_{i=0}^{N_b-1} \frac{\tilde{b}_i}{\lambda_i}|\lambda_i\rangle_b\otimes |N_c t\frac{ \lambda_i}{2\pi}\rangle_c\otimes
|1\rangle_a \,.
 \end{align}
 We can use the normalization $\langle\tilde x|\tilde x\rangle=1$ to write $\sum_l | b_l/\lambda_l |^2=1$.
However, we cannot write the $b$-block register as the $|x\rangle_b$ state yet since it is entangled with the $c$-block qubits indicated by the sum on the index $i$. 

The third step in HHL applies an inverse QPE that disentangles the $b$- and $c$-block registers. This works because inverse QPE (or QPE in reverse order) involves quantum Fourier transformation which results in $|N_c t\frac{ \lambda_i}{2\pi}\rangle_c=\sum_y\exp[iy \lambda_i t]|y\rangle/\sqrt{N_c}$, followed by application of a series of $U^\dagger(t)$ controlled on the $c$-block state $|y\rangle$ resulting in a phase kick-back that exactly cancels the exponential. The state comprised of the $b$-block, $c$-block and flag-qubit registers become
\begin{multline}
\sum_{i=0}^{N_b-1} \frac{\tilde{b}_i}{\lambda_i}|\lambda_i\rangle_b\otimes \frac{1}{\sqrt{N_c}}\sum_{y=0}^{N_c-1}|y\rangle_c\otimes
|1\rangle_a\\ = |\tilde x\rangle \otimes \frac{1}{\sqrt{N_c}}\sum_{y=0}^{N_c-1}|y\rangle_c\otimes
|1\rangle_a\,.
\end{multline}
It is customary to apply Hadamard gates to the $c$-block qubits to return them to the $|0\rangle_c$ state. 
A measurement of the $b$-block qubit register yields the components of the vector $\vec{\bm x}$ as probabilities for $(|x_r|^2, |\vec{\bm x}_+|^2, |x_s|^2, |\vec{\bm x}_-|^2)$. This is sufficient to calculate
\begin{align}
\mathcal T(E)=|s|^2=c^2{|x_s|^2}= \frac{|x_s|^2}{ |x_r|^2}\,.
\end{align}

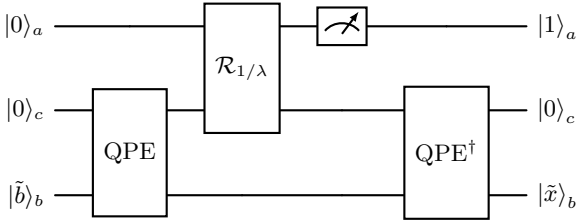
\begin{figure}[htb]
\begin{center}
\begin{quantikz}
\lstick{$|0\rangle_a$}& &  \gate[wires=2]{\mathcal{R}_{1/\lambda}} &  \meter{}  &     & \rstick{$ \ket{1}_a$}\\
\lstick{$|0\rangle_c$}& \gate[wires=2]{\mathrm{QPE}} &&  &\gate[wires=2]{\mathrm{QPE}^\dagger} & \rstick{$ \ket{0}_c$}\\
\lstick{$|\tilde b\rangle_b$}  & & & &  & \rstick{$ \ket{\tilde x}_b$}
\end{quantikz}
\end{center}
\caption{Modular circuit for HHL. The $b$-block (system) register is initialised to the state $|\tilde b\rangle_b=\ket{0}_b$ due to the rescaling of the original $\vec{\bm b}$ in Eq.~(\ref{eq:QLSE}). QPE entangles the clock register $\ket{0}_c$ with the eigenphases of $\exp({i\tilde{A}t})$. The block $\mathcal{R}_{1/\lambda}$ implements eigenvalue inversion on the clock register conditioned on the $b$-block register. Proceeding when the flag qubit measures $|1\rangle_a$, inverse QPE uncomputes the clock register, and projects the $b$-block register  onto  the solution state $|\tilde x\rangle_b$.}
\label{fig:hhl_circuit}
\end{figure}

In Fig.~\ref{fig:hhl_circuit}, we show the entire HHL circuit as described above.  

\section{Variational Quantum Linear Solver}
\label{sec:VQLS}

The second quantum algorithm that we use for calculating $\mathcal{T}(E)$ is VQLS~\cite{BravoPrieto:2023} which is a variational method. 
In VQLS, the variational states are parameterized as  $|\alpha\rangle=(\alpha_r, \vec{\bm \alpha}_+,\alpha_s, \vec{\bm \alpha}_-)$. This yields a normalization constant $c=(\alpha_r+\alpha_r^\ast)/(|\alpha_r|^2+|\alpha_s|^2)$. 
Thus, it would seem we need to calculate $c$ for every variational state and rescale the matrix $\xbar{A}$ as we run our iterative calculation. However, we see below in Eqs.~(\ref{eq:globalcost}), (\ref{eq:localcost}) that in the variational calculation where we work with normalized states, the rescaling of the matrix $A$ with $c$, which can be done classically, is not necessary.

We can understand the VQLS algorithm as follows. Suppose we construct a 
variational state $|\bm{\alpha}\rangle=V(\bm{\alpha})|0\rangle$ using a unitary operator $V(\bm{\alpha})$. The goal is to optimize $\bm{\alpha}$ such that $A|\bm\alpha_0\rangle=e^{-i\theta}|b\rangle$ which is the solution $|x\rangle=|\bm\alpha_0\rangle$ we seek up to a phase $\theta$. Alternatively, for $|\psi(\bm\alpha)\rangle = A|\bm{\alpha}\rangle$, we seek to minimize the  overlap of $|\psi(\bm\alpha)\rangle$ with the component orthogonal to $|b\rangle$. Thus we look at the cost function~\cite{BravoPrieto:2023}
\begin{align}\label{eq:globalcost}
   C(\bm\alpha)= \frac{\langle\psi(\bm\alpha)|\left(\mathbbm{1}-|b\rangle\langle b|\right)|\psi(\bm\alpha)\rangle}{\langle \psi(\bm\alpha)|\psi(\bm\alpha)\rangle}\,.
\end{align}
We see that in the computation of the cost function $C(\bm\alpha)$, the normalization $c$ can be eliminated without affecting the variational method. 
On the quantum computer if we represent $|b\rangle=U|0\rangle$ with an unitary matrix $U$, then the cost function can be written as an expectation value of a Hamiltonian:
\begin{align}
    C(\bm\alpha) &=  \frac{\langle\bm\alpha|H|\bm\alpha\rangle}{\langle \psi(\bm\alpha)|\psi(\bm\alpha)\rangle}\,,\nonumber\\
    H&=\xbar{A}^\dagger U \left(\mathbbm{1}-|0\rangle\langle 0|\right)U^\dagger \xbar{A}\,.
\end{align}

The authors in Ref.~\cite{BravoPrieto:2023} suggest an alternate cost function
\begin{align}\label{eq:localcost}
    C_L(\bm\alpha) &=  \frac{\langle\bm\alpha|H_L|\bm\alpha\rangle}{\langle \psi(\bm\alpha)|\psi(\bm\alpha)\rangle}= 
    \frac{\langle\bm\alpha|H_L|\bm\alpha\rangle}{\langle\bm\alpha|\xbar{A}^\dagger \xbar{A}|\bm\alpha\rangle}
    \,,\nonumber\\
    H_L&=\xbar{A}^\dagger U \left(\mathbbm{1}-\sum_{j=0}^{N-1}|0_j\rangle\langle 0_j|\otimes\mathbbm{1}_{\bar{j}}\right)U^\dagger \xbar{A}\nonumber\\
    &=\xbar{A}^\dagger U \left(\frac{1}{2}\mathbbm{1}-\frac{1}{2N}\sum_{j=0}^{N-1} Z_j \right) U^\dagger \xbar{A}\,,
\end{align}
where $|0_j\rangle$ is a zero state on the $j$-th qubit and $\mathbbm{1}_{\bar j}$ is an identity on all qubits bar the $j$-th qubit. 
The local cost function $C_L(\bm\alpha)$ is claimed to avoid barren plateaus where the cost functions develop flat directions with vanishing gradients as the number of qubits $N$ increases~\cite{Cerezo:2021}. A useful bound exists that implies that $C_L(\bm\alpha)\leftrightarrow C(\bm\alpha)=0$ i.e. minimization of $C_L(\bm\alpha)$ would also minimize $C(\bm\alpha)$ that would result in the solution $|x\rangle=|\bm\alpha_0\rangle$. For future reference, we will notate $\ket{x_\text{VQLS}}=|\bm\alpha_0\rangle$.

Note that $|0_j\rangle\langle 0_j|=(\mathbbm{1}_j-Z_j)/2$ results in a single Pauli matrix $Z_j$ instead of the original $|0\rangle\langle0|$ that results in Pauli strings with up to as many $Z$s as the number of qubits. The expectation values are calculated using the Hadamard test so shorter Pauli strings results in fewer CNOT gates. 

The quantum computation of the local cost function $C_L(\bm\alpha)$ involves decomposition of the matrix $\xbar{A}$ (which may be non-Hermitian besides being non-unitary)
\begin{align}
\xbar{A}=\sum_{l=0}^{L-1} c_l A_l\,,
\end{align}
where $A_l$s are tensor products of Pauli matrices defined on the $N$-qubit tensor product space. $N=\log_2(\text{dim})$ where $\text{dim}$ is the dimension of $\xbar{A}$. Then the normalization of the wavefunction becomes
\begin{align}
   \langle\psi(\bm\alpha)|\psi(\bm\alpha)\rangle&=\langle\bm\alpha |\xbar{A}^\dagger \xbar{A}|\bm\alpha\rangle=\sum_{m,n=0}^{L-1} c_m^\ast c_n\beta_{mn}\,,\nonumber\\
   \beta_{mn}&=\langle\bm\alpha|A_m^\dagger A_n|\bm\alpha\rangle\,,
\end{align}
and the expectation of the Hamiltonian $H_L$
\begin{align}
\langle\bm\alpha|H_L|\bm\alpha\rangle&=\frac{\langle\bm\alpha|\xbar{A}^\dagger \xbar{A}|\bm\alpha\rangle}{2} -\sum_{j=0}^{N-1}\sum_{m,n=0}^{L-1}c_m^\ast c_n \frac{\Delta_{mn}^{(j)}}{2N}\,,\nonumber\\
\Delta_{mn}^{(j)}&= \langle\bm\alpha|A_m^\dagger U Z_j U^\dagger A_n|\bm\alpha\rangle\,.
\end{align}

\begin{figure}[htb]
\begin{center}
\begin{quantikz}
\lstick{$|0\rangle$}&\gate{H} & \gate[style={fill=red!20},label style=black]{S^\dagger}&\ctrl{1}& \gate{H} &\meter{}\\
\lstick{$|\bm\alpha\rangle$}  &\qwbundle{n} & &\gate{M} & &
\end{quantikz}
\end{center}
\caption{The Hadamard test circuit for calculating $\langle\bm\alpha|M|\bm\alpha\rangle=\langle M\rangle_{\bm\alpha}$. The phase gate $S^\dagger$ is needed to calculate the imaginary component of $\langle M\rangle_{\bm\alpha}$.}
\label{fig:Hadamard}
\end{figure}
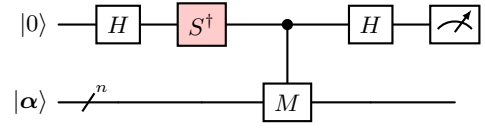

The calculations of both $\beta_{mn}$ and $\Delta_{mn}^{(j)}$ are done by the Hadamard test, Fig.~\ref{fig:Hadamard}. The expectation value of a Pauli string $M$ in the state  $|\bm\alpha\rangle$, $\langle M\rangle_{\bm\alpha}\equiv \langle\bm\alpha|M|\bm\alpha\rangle$, is obtained from the measurement of the ancilla qubit in  the circuit in Fig.~\ref{fig:Hadamard} as shown. 
In particular, measurements in the circuit without the phase gate $S^\dagger$ gives 
 $\operatorname{Re}[\langle M\rangle_{\bm\alpha}]=P_0(\bm\alpha)-P_1(\bm\alpha)$ where $P_0(\bm\alpha)$ and $P_1(\bm\alpha)$ are the probabilities of measuring the ancilla qubit in the state $|0\rangle $ or $|1\rangle$, respectively, when the system is in the state $|\bm\alpha\rangle$. Similarly, the imaginary part is obtained from measurements in the circuit with the phase gate $S^\dagger$ as $\operatorname{Im}[\langle M\rangle_{\bm\alpha}]=P_0(\bm\alpha)-P_1(\bm\alpha)$. 

The Pauli strings $A_m$s are Hermitian as they are a tensor products of Pauli matrices or the identity matrix $\mathbbm{1}_{2\times2}$. Thus the diagonal elements of $\beta_{mn}$ and $\Delta_{mn}^{(j)}$ are real, and $\beta_{nm}=\beta_{mn}^\ast$, 
$\Delta_{nm}^{(j)}=[\Delta_{mn}^{(j)}]^\ast$ which reduces the number of required Hadamard test measurements. Further, $\beta_{mm}=1$ and we do not calculate these elements directly. Thus we require $2\times L(L-1)/2=L(L-1)$ measurements for $\beta_{mn}$. In the $\Delta_{mn}^{(j)}$ computations, the diagonal elements require $L$ measurements and the off-diagonal elements require $2\times L(L-1)/2=L(L-1)$ measurements for each $j$. Thus the total measurements would amount to $(1+N)L^2-L$.  Moreover, a property of the Hadamard test is that when $M= A_m^\dagger Z_j A_m$ for the diagonals of $\Delta_{mn}^{(j)}$, only a CNOT gate on the $Z_j$ operator is required simplifying the calculation.  

Note that due to the rescaling of the original vector $b$ in the linear system of equations, the unitary matrix $U=\mathbbm{1}$ in our circuits.

We supplement the local cost function $C_L(\bm\alpha)$ calculation by a physical constraint for perfect transmission $|r|^2=|s|^2$ which requires the  $\alpha_r$ and the $\alpha_s$ component of the trial state $|{\bm \alpha}\rangle$ to have the same magnitude. We find solutions that satisfy the perfect transmission $\mathcal T(E)=1$ condition. 

\subsection{Trial State Preparation for VQLS}
\label{sec:trialstate}

The VQLS computations involve variational calculation with either a 2-qubit state or a 3-qubit state which has the generic forms:
\begin{multline}\label{eq:generic_states}
|\bm\alpha\rangle_\text{2-qubit}= \alpha_0|00\rangle+\alpha_1|01\rangle+\alpha_2|10\rangle+\alpha_3|11\rangle\,,\\
|\bm\alpha\rangle_\text{3-qubit}=\alpha_0|000\rangle+\alpha_1|001\rangle+\alpha_2|010\rangle+\alpha_3|011\rangle\\
+\alpha_4|100\rangle+\alpha_5|101\rangle+\alpha_6|110\rangle+\alpha_7|111\rangle\,.
\end{multline}
The probability amplitudes $\alpha_i$s are complex, however, we can choose an overall phase for $\alpha_0$ to be real and also require the states to have $l^2$-norm equal to 1. Thus we can prepare a generic 2-qubit and 3-qubit states with 7 and 15 variables, respectively, with another reduction coming from the normalization. 

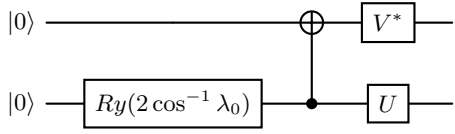
\begin{figure}[htb]
\begin{center}
\begin{quantikz}
\lstick{$|0\rangle$}& &\targ{}&\gate{V^\ast}&\\
\lstick{$|0\rangle$}  &\gate{Ry(2\cos^{-1}\lambda_0)}&\ctrl{-1}&\gate{U}&
\end{quantikz}
\end{center}
\caption{ Circuit for a generic 2-qubit state as described in the text.}
\label{fig:2qubit}
\end{figure}

The discussion here is based on the 2-qubit state preparation from Ref.~\cite{Yusf:2024igb}. First we see that an arbitrary 1-qubit state $a|0\rangle+b|1\rangle$ for real $a$ can be prepared using unitary gates 
on the initial state $|0\rangle$. To construct a generic 2-qubit state from Eq.~(\ref{eq:generic_states}), we first bipartite the Hilbert space 
\begin{align}
|\bm\alpha\rangle_\text{2-qubit}&= \sum_{a,b=0}^1 M_{ab}|a\rangle\otimes|b\rangle\,,\nonumber\\
M&=\begin{pmatrix} \alpha_0&\alpha_1\\ \alpha_2&\alpha_3
\end{pmatrix}\,.
\end{align}
Singular value decomposition (SVD) gives $M=USV^\dagger$ where $S$ is a diagonal $\operatorname{diag}(\lambda_0,\lambda_1)$ matrix with $\lambda_0\geq\lambda_1\geq0$. Schmidt decomposition in terms of new orthonormal basis $|i\rangle_A=\sum_{a}U_{ai}|a\rangle$, $|i\rangle_B=\sum_{b}V^\ast_{bi}|b\rangle$ allows one to write
\begin{multline}
|\bm\alpha\rangle_\text{2-qubit}=\sum_{i=0}^1\lambda_i|i\rangle_A \otimes|i\rangle_B\\
= (U\otimes V^\ast)(\lambda_0|00\rangle+\lambda_1|11\rangle)\,.
\end{multline}
Then the desired 2-qubit state can be prepared from initial state $|00\rangle=|0\rangle\otimes|0\rangle$ by first creating a superposition state $(\lambda_0|0\rangle+\lambda_1|1\rangle)\otimes|0\rangle$ with the single rotation $Ry(2\cos^{-1}\lambda_0)$ on the higher qubit since both $\lambda_0$ and $\lambda_1$ are real. Then a CNOT gate with the control on the higher qubit targeting the lower qubit generates the entangled state $\lambda_0|00\rangle+\lambda_1|11\rangle$. Finally, single gate operations apply the unitary matrices $U$, $V^\ast$ on the higher and lower qubits, respectively to produce the desired states. The entire process is depicted in the circuit in Fig.~\ref{fig:2qubit}.

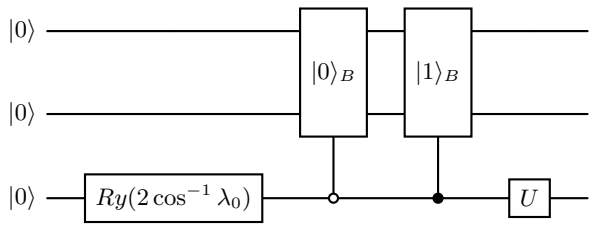
\begin{figure}[htb]
\begin{center}
\begin{quantikz}
\lstick{$|0\rangle$}& &\gate[2]{|0\rangle_B}& \gate[2]{|1\rangle_B}& &\\
\lstick{$|0\rangle$}& & & &&\\
\lstick{$|0\rangle$}  &\gate{Ry(2\cos^{-1}\lambda_0)}&\octrl{-1}&\ctrl{-1}&\gate{U}&
\end{quantikz}
\end{center}
\caption{ Circuit for a generic 3-qubit state as described in the text. The first controlled unitary operation creates the $|0\rangle_B$ state on the lowest two qubits if the highest qubit is in state $|0\rangle$, and the second controlled unitary operation creates the $|1\rangle_B$ state on the lowest two qubits if the highest qubit is in state $|1\rangle$.}
\label{fig:3qubit}
\end{figure}

The generic 3-qubit state $|\bm\alpha\rangle_\text{3-qubit}$ from Eq.~(\ref{eq:generic_states}) is constructed similar to the 2-qubit state $|\bm\alpha\rangle_\text{2-qubit}$ through a divide-and-conquer strategy. First we divide the Hilber space into subspace $A$ of the highest qubit, and subspace $B$ of the lowest qubits such that
\begin{align}
|\bm\alpha\rangle_\text{3-qubit}&= \sum_{a=0}^1 \sum_{b=0}^3 M_{ab}|a\rangle\otimes|b\rangle\,,\nonumber\\
M&=\begin{pmatrix} \alpha_0&\alpha_1& \alpha_2&\alpha_3\\ \alpha_4&\alpha_5& \alpha_6&\alpha_7
\end{pmatrix}\,=U\begin{pmatrix}\lambda_0&0\\
0&\lambda_1\end{pmatrix}V^\dagger,
\end{align}
which gives 
\begin{align}
|\bm\alpha\rangle_\text{3-qubit} = \lambda_0|0\rangle_A \otimes|0\rangle_B+\lambda_1|1\rangle_A\otimes|1\rangle_B\,,
\end{align}
with orthonormal basis $|i\rangle_A=\sum_{a}U_{ai}|a\rangle$ as before, and also 2-qubit orthonormal basis 
$|i\rangle_B=\sum_{b}V_{bi}^\ast|b\rangle$. The circuit in Fig.~\ref{fig:2qubit} can be used to create the 2-qubit  states $|i\rangle_B$. So, to create a generic 3-qubit state we first create the state $(\lambda_0|0\rangle+\lambda_1|1\rangle)\otimes |00\rangle$ with a  $Ry(2\cos^{-1}\lambda_0)$ on the highest qubit. Then we create the superposition state $\lambda_0|0\rangle\otimes|0\rangle_B+\lambda_1|1\rangle\otimes|1\rangle_B$ by conditioning on the highest qubit as depicted in Fig.~\ref{fig:3qubit}. The final $|\bm\alpha\rangle_\text{3-qubit}$ is produced by applying the $U$ operator on the highest qubit.

\section{Results and Analysis}
\label{sec:results}

In all the numerical work, we take the 
hopping strength $t_L=\SI{2.7}{\eV}$ and lead onsite energy $\epsilon_L=\SI{-2.7}{\eV}$. These parameters are motivated by the electronic structure of metallic armchair SWCNTs, which serve as canonical physical realizations of quantum dragon nanodevices with perfect ballistic transmission~\cite{NOVO2014, Inkoom2018}. The nearest-neighbor carbon--carbon tight-binding hopping integral in graphene and armchair SWCNTs is well established in the range $2.7$--$2.8$~eV~\cite{CastroNeto2009, Reich2002, Saito1992}, and our choice of $t_L = 2.7$~eV reflects this value directly. When armchair SWCNTs serve as semi-infinite leads connected to a nanodevice, the hopping strength within the leads equals the intralayer carbon--carbon bond hopping~\cite{NOVO2014}, making $t_L$ a material parameter fixed by experiment rather than a free fitting constant. The lead onsite energy $\epsilon_L = -t_L = -2.7$~eV is chosen so that the charge-neutrality (Dirac) point of graphene, conventionally placed at $E = 0$, falls within the conducting window $[\epsilon_L - 2t_L,\; \epsilon_L + 2t_L] = [-8.1,\; 2.7]$~eV. This ensures that the energy range explored in our calculations encompasses the Fermi-level regime most relevant to room-temperature ballistic transport experiments on carbon nanotube and graphene junctions~\cite{Frank1998, Bolotin2008, javey2003ballistic}. In particular, setting $\epsilon_L = -t_L$ guarantees that the dimensionless scaled energy
\begin{align}
    E_{\mathrm{scaled}} = \frac{2 (E-E_{\mathrm{min}})}{E_{\mathrm{max}}-E_{\mathrm{min}}} -1=\frac{E-\epsilon_L}{2 t_L}\,,
    \label{eq:escaled}
\end{align}
maps the full conducting bandwidth onto the symmetric interval $[-1, 1]$, placing the zero-energy Fermi level at $E_{\mathrm{scaled}} = 0.5$ and centering the band-edge divergences of the condition number symmetrically at $E_{\mathrm{scaled}} = \pm 1$. Together, these parameters constitute a realistic and experimentally grounded model for electron transport through carbon nanotube junctions, providing a physically meaningful benchmark for the quantum computational methods developed here.

For the quantum computation, we work with the systems defined in Eqs.~(\ref{eq:2siteDragonAxb}), (\ref{eq:6siteDragonAxb}) for the 2-site and 6-site quantum dragon devices, respectively, obtained after similarity transformation as detailed earlier.

\subsection{2-site Quantum Dragon with HHL}
\label{subsec:2sitehhl}

The Hermitian $8\times 8$ matrix $\tilde A$ for the 2-site device in Eq.~(\ref{eq:Atilde}) is block-encoded with $n_b = 3$ qubits on the quantum register. The HHL algorithm has two parameters---number of clock qubits $n_c$ and time parameter $t$. The use of \verb|ExactReciprocalGate()| introduces the additional scaling parameter $s$ in the computation.  
These parameters control different aspects of the HHL algorithms. Typically larger $n_c$ gives more precise determination of the spectrum $\lambda_i$. However, one has to consider accuracy as well. This is where the spectrum $\lambda_i$ and the gap in the spectrum play a role in choosing a suitable time evolution $t$ since phase kickback involves factors of $\lambda_i t$.  Finally, the scaling factor $s$ controls the success probability of measuring the flag qubit in state $\ket{1}_a$ by a factor $2\pi s/(t\lambda_i)$ from Eqs.~(\ref{eq:grover}), (\ref{eq:exactreciprocal}). A natural definition in terms of angular frequency $2\pi/t$ leads one to consider
\begin{align}
    s= s' \frac{t}{2\pi}\,.
\end{align}
One can pick some nominal values $t=\SI{1}{\eV^{-1}}$, $s'=\SI{0.5}{\eV}$ (keeping probabilities much less than 1) and $n_c=6$ (for about $2^{-n_c}\sim 0.02$ accuracy).

\begin{figure}[htb]
\begin{center}
\includegraphics[width=0.49\textwidth]{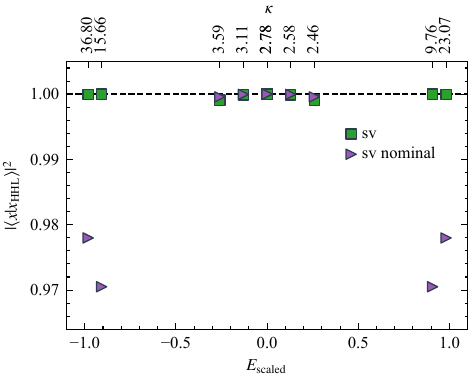}
\caption{ Exact statevector calculation of $|\langle x_\text{HHL}|\tilde{x}\rangle|^2$. The bottom horizontal axis show   rescaled energy $E_\text{scaled}$ and the top horizontal axis shows the condition number $\kappa$ for $\tilde A$ from Eq.~(\ref{eq:Atilde}).  The (green) square data points are statevector calculations with the tuned parameters from Table~\ref{table:HHLparams}, and the (purple) triangle data points are with nominal parameters  described in the text.}
\label{fig:HHLparams}
\end{center}
\end{figure}

A ``statevector" calculation where the HHL circuit is written as the action of an unitary matrix acting on the initial state $\ket{0}_b\otimes\ket{0}_c\otimes\ket{0}_a$ comprised of the physical block qubits, counting clock qubits and the ancilla flag qubit, as described in Section~\ref{sec:hhl},  can be used to set some benchmark estimates. In Fig.~\ref{fig:HHLparams}, we show the exact numerical calculation of the overlap $|\langle x_\text{HHL}|\tilde{x}\rangle|^2$ where $\ket{x_\text{HHL}}$ is the statevector calculation in Qiskit~\cite{Qiskit}. The exact result $\ket{\tilde x}$ is obtained from the classical numerical solution to the LSE in Eq.~(\ref{eq:Atilde}). Specific parameter choices, for example, $\beta=23^\circ$, and $t_{ab}=t_L=\SI{2.7}{\eV}$, allows one to cast the dragon solution to the form relevant to the original LSE in Eq.~(\ref{eq:2siteNonLinearDevice})  as discussed in subsection~\ref{subsec:2siteDragon}.

We see in Fig.~\ref{fig:HHLparams} that the nominal choices $t=\SI{1}{\eV^{-1}}$, $s'=0.5$, $n_c=6$ gets within a few percent of the expected result $|\langle x_\text{HHL}|\tilde{x}\rangle|^2=1$. With a little tuning, which is possible in this case since we know the exact solution $\ket{\tilde x}$, with parameters listed in Table~\ref{table:HHLparams} we get to well within a fraction of a percent of the true result. In Table~\ref{table:HHLparams} , we also provide the exact statevector calculation of the success probability $P_\text{success}$, defined as the probability of measuring the flag qubit in the state $\ket{1}_a$. We also list the statevector calculation of transmission coefficient $\mathcal T$ which shows near perfect transmission of 1. In the rest of the HHL calculations, we use the tuned parameters. 
\begin{table}[htb]
\centering
\caption{\protect Parameters $t$ and $n_c$ for the HHL algorithm with a fixed $s'=\SI{0.5}{\eV}$. Results for the statevector calculations of $P_\text{success}$, $\langle\tilde x|x_\text{HHL}\rangle$ and $\mathcal T(E)$ are listed.}
\begin{ruledtabular}
\begin{tabular}{rccccc}
$E$ (\si{\eV})  & $t (\si{\eV^{-1}})$ 
  &$n_c$ & $ P^{(\text{exact})}_\text{success}$ (\%) &  $|\langle \tilde x\ket{x_\text{HHL}}|^2$ & $\mathcal T_\text{exact}$
\\ \hline \rule{0pt}{0.9\normalbaselineskip}
\begin{filecontents*}{HHL_params.csv}
energy,escaled,time,scaleprime,precision,pexact,oexact,texact
-8,-0.981,0.3,0.5,6,19.69,1,1.001
-7.6,-0.907,0.6,0.5,6,19.61,1,1.003
-4.1,-0.259,0.3,0.5,7,18.86,0.999,0.993
-3.4,-0.13,0.664,0.5,6,17.92,1,1.008
-2.701,-1.90E-04,0.6,0.5,6,19.33,1,1.001
-2.699,1.90E-04,0.6,0.5,6,19.33,1,1.001
-2,0.13,0.664,0.5,6,17.92,1,1.008
-1.3,0.259,0.3,0.5,7,18.86,0.999,0.993
2.2,0.907,0.6,0.5,6,19.61,1,1.003
2.6,0.981,0.3,0.5,6,19.69,1,1.001
\end{filecontents*}
\csvreader[head to column names, late after line=\\]{HHL_params.csv}{}
{\ \num{\energy} 
&  \num{\time} &\num{\precision}
&\pexact &\num{\oexact} &\num{\texact}

}
\end{tabular}
\end{ruledtabular}
 \label{table:HHLparams}
\end{table}

We provide a general strategy for tuning $t$ when the exact dragon solution is not known apriori. We saw earlier that in the dragon solution, Eqs.~(\ref{eq:2siteDragonAxb}), (\ref{eq:6siteDragonAxb}), the wavefunction for the leads and the component $\psi_+$ and $\psi_{j,+}$ for the 2-site and the  6-site devices, respectively, can be represented as an equivalent linear device independent of the null wavefunctions $\psi_-=0$ and $\psi_{j,-}=0$. In the equivalent linear device with perfect transmission, the magnitudes of the wavefunctions at each site are equal.  For the 2-site device with only 3 non-zero amplitudes of equal magnitude (two leads plus $\psi_+$), normalization of the wavefunction implies the non-zero amplitudes have magnitude $1/\sqrt{3}$. Similarly,  for the 6-site device, we expect 5 non-zero amplitudes of magnitude $1/\sqrt{5}$. As per the formulation of the 2-site state $\ket{\tilde x}$ in Eq.~(\ref{eq:Atilde}), measurement of the 8-dimensional block-qubit state in the HHL algorithm should return an 8-dimensional probability vector with the \qtyrange{5}{7}{th} component being ideally $1/3$.  Here we tuned $t$ from measurements in ideal simulations. On real quantum hardware, it would be an expensive exercise but a valid strategy. In Appendix~\ref{appendix:tuning_t}, we discuss tuning of $t$ in more detail. Alternatively, the number, location, and size of the non-zero components of the block qubit states from its probability measurements could be a necessary but not sufficient check if the measured state is compatible with the dragon solution.

\begin{figure}[htb]
\begin{center}
\includegraphics[width=0.49\textwidth]{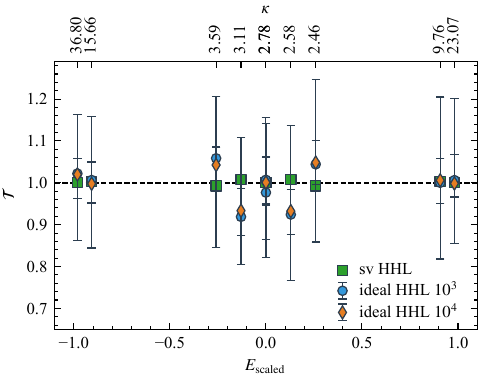}
\caption{ Transmission coefficient $\mathcal T(E)$ vs energy $E_\text{scaled}$ and condition number $\kappa$. The (green) square data points are from statevector calculations, (blue) circle data points for ideal simulations with $10^3$ shots, and (orange) diamond data points for ideal simulations with $10^4$ shots. The HHL parameters are from Table~\ref{table:HHLparams}.}
\label{fig:HHLtransmission}
\end{center}
\end{figure}

The exact $P_\mathrm{success}^\text{(exact)}$ from the statevector calculation varies $\sim20\%$ across the scattering energy band, Table~\ref{table:HHLparams}. Thus, in the ideal simulations, the number of shots $\numrange{1000}{10000}$ would result in about $\qtyrange{7}{2}{\%}$ uncertainty from the finite number of shots. We perform 100 independent measurements with both $10^3$ and $10^4$ shots at each energy to obtain some statistics. The quoted errors in the ideal and later in the noisy simulations are derived from the 100 distributions at each energy. One can add an additional $\sim\qtyrange{7}{2}{\%}$ uncertainty  from the finite shots as well as appropriate. The ideal simulation results for transmission coefficient $\mathcal T(E)$ are shown in Fig.~\ref{fig:HHLtransmission}. The ideal simulation results are encouraging. They accurately calculate the expected result within the uncertainty across the energy band except at two energies mid-band. However, once the uncertainty from the finite shots are added, the exact results are reproduced.  

\begin{figure}[htb]
\begin{center}
\includegraphics[width=0.49\textwidth]{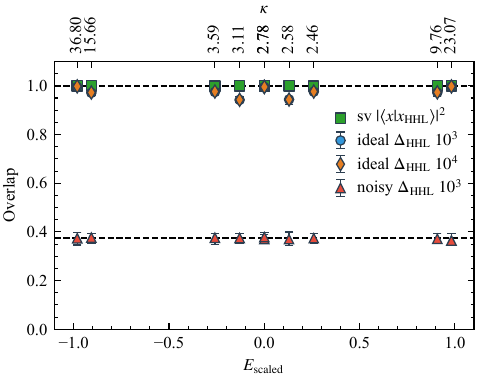}
\caption{ 
The (green) square data points statevector overlap $|\langle x_\text{HHL}\ket{x}|^2$, (blue) circle ideal simulation overlap $\Delta_\text{HHL}$ with $10^3$ shots, (orange) diamond ideal simulation overlap $\Delta_\text{HHL}$ with $10^4$ shots, and 
(red) triangle noisy \texttt{Fake\_Torino} simulation overlap $\Delta_\text{HHL}$ with $10^3$ shots. The horizontal dashed grid line is at 0.375. Rest of the information is the same as Fig.~\ref{fig:HHLtransmission}. }
\label{fig:HHLoverlap}
\end{center}
\end{figure}

Next we consider how accurately, the state $\ket{\tilde x}$ is reproduced in the simulation.
In the HHL algorithm, we do not have access to the solution state $\ket{x_\text{HHL}}$. Instead we only have information about the absolute-value square of the amplitudes $|x_i|^2$ of the wavefunction $\ket{x_\text{HHL}}$,  from the measurements as probabilities,  where $x_i$ are the components (amplitudes) of $\ket{x_\text{HHL}}$. Suppose we construct a vector $|\sqrt{x_\text{HHL}^2}\rangle$ with $|x_i|$ as the components which contains no information about the relative phases between the amplitudes in the state $\ket{x_\text{HHL}}$. Then a \emph{necessary} requirement on the quality of the solution obtained from HHL would be the overlap 
$\Delta_\text{HHL}\equiv|\langle\sqrt{\tilde{x}^2}|\sqrt{x_\text{HHL}^2}\rangle|^2$ where we constructed the vector $|\sqrt{\tilde{x}^2}\rangle$ from the magnitude of the exact solution $\ket{\tilde x}$ amplitudes. We \emph{emphasize} that $\Delta_\text{HHL}=1$ does not guarantee a perfect solution since all relative phase information  in $\ket{\tilde x}$ are ignored in  $\Delta_\text{HHL}$. However, a $\Delta_\text{HHL}$ much smaller than 1 necessarily indicate that the HHL algorithm didn't calculate the exact solution accurately. With this \emph{caveat}, we plot $\Delta_\text{HHL}$ from the ideal simulations in Fig.~\ref{fig:HHLoverlap} along with the statevector result for $|\langle x_\text{HHL}\ket{x}|^2$. 

\begin{figure}[htb]
\begin{center}
\includegraphics[width=0.49\textwidth]{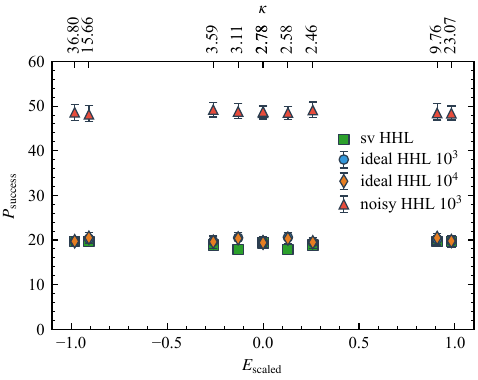}
\caption{ Success probabilities $P_\text{success}$ as described in the text with simulation parameters from Figs.~\ref{fig:HHLtransmission}, \ref{fig:HHLoverlap}.}
\label{fig:HHLprobabilities}
\end{center}
\end{figure}

We see in Fig.~\ref{fig:HHLoverlap}, $\Delta_\text{HHL}\approx1$ for the two ideal simulations across the entire energy band. This does not prove that the measurement results are from the true state $\ket{\tilde x}$ we are seeking since the $\Delta_\text{HHL}$ from the measurement probabilities all relative phase information is lost. In practice, to obtain the wavefunction $\ket{x_\text{HHL}}$, one can perform a subsequent Hadamard test with added computational overhead to obtain the phase information. In the simulation, we do a further check to confirm that the measurement of the physical block qubits $\ket{\tilde x}_b$ returns 8-dimensional vector of probabilities, for the 2-site device with the first four elements being zero in agreement with Eq.~(\ref{eq:Atilde}).  We also calculated the success probabilities for the ideal simulations. These are shown in Fig~\ref{fig:HHLprobabilities}. They are in agreement with the exact statevector values.

Results from noisy simulations with a noise model based on \verb|ibm_torino| in Qiskit~\cite{Qiskit} are shown in Figs.~\ref{fig:HHLoverlap} and \ref{fig:HHLprobabilities}. These results were obtained from 1000 shots with 100 samples at each energy $E$ (or $E_\text{scaled}$) like the ideal simulations.  In the noisy simulations, the overlap $\Delta_\text{HHL}\ll1$, averaged over 100 independent experiments at each $E_\text{scaled}$,  across the transmission energy band. This indicates poor overlap of the HHL result with the exact state $\ket{\tilde x}$. At first glance, the noisy simulation overlap of about $\Delta_\text{HHL}\approx 0.4$ uniformly across the entire energy band seem suspicious. However, this value can be understood precisely. If one looks at the $P_\text{success}$ for the noisy simulations, one sees that the success probabilities are much higher than the statevector estimates. The reason behind this is that the error model calculations results in many false positives for the $\ket{1}_a$ flag qubit state leading to large success rates. Consequently, all the probabilities in the 8-component block qubit state are just random variables with equal probabilities of 1/8. This leads to a $|\sqrt{x_\text{HHL}^2}\rangle$ whose all 8 components are $1/\sqrt{8}$. On the other hand $|\sqrt{\tilde{x}^2}\rangle$ across the energy band has only three non-zero components of equal size $1/\sqrt{3}$. 
 One then finds that $\Delta_\text{HHL}\equiv|\langle\sqrt{\tilde{x}^2}|\sqrt{x_\text{HHL}^2}\rangle|^2=(3/\sqrt{24})^2\approx0.375$ deterministically because the noise model produces the measured probabilities uniformly without any structure. At the same time, the noisy model will produce a perfect transmission $\mathcal T(E)=1$ since it is calculated from ratio of probabilities that are all exactly $1/8$.

We observe that the accuracy in the ideal simulations seem unaffected across the energy band even as the condition number $\kappa$ for the matrix $\tilde{A}$ changes by an order of magnitude, see Figs.~\ref{fig:HHLtransmission}, \ref{fig:HHLoverlap}, ~\ref{fig:HHLprobabilities}. However, the noisy simulation results led us to consider a variational approach to solving for the dragon solution. The deep QPE circuits in HHL makes it an impractical approach on NISQ devices. We next present results from the VQLS algorithm.

\subsection{2-site Quantum Dragon with VQLS}
\label{subsec:results_2site}
In the VQLS implementation we make a small modification to the Eqs.~(\ref{eq:2siteDragonAxb}), (\ref{eq:6siteDragonAxb}) for 2-site and 6-site devices without affecting the final result. 
 The dragon solution have $|\bm\Psi_{-}\rangle=|0\rangle$. Thus, the solution with vector components $(1+r, |\bm\Psi_{+}\rangle, s, |\bm\Psi_{-}\rangle)$ is independent of the matrix sub-block spanned by the $|\bm\Psi_{-}\rangle$. Therefore, we replace this block with diagonal elements $\epsilon_L-E$ which reduces the number of Pauli string $L$ needed for computation. For the 2-site nonlinear device $L=6$ instead of $L=8$
 whereas for the 6-site device $L=14$ instead of $L=18$, respectively. This reduces the number of circuits from 184 to 102 for the 2-site device , and from 1278 to 770 for the 6-site device. A smaller set of Pauli strings results in greater accuracy. We can recover the wavefunctions at each slice of the original matrices in Eqs.~(\ref{eq:2siteNonLinearDevice}), (\ref{Eq:6siteDragon:01})  from $\psi_{1,j}=\psi_{j,+}\cos\beta_j$, $\psi_{2,j}=\psi_{j,+}\sin\beta_j$ for arbitrary $0<\beta_j<\pi/2$. 

 In principle the same modification as above could have been done for the HHL algorithm. However, in the HHL simulations we constructed the unitary operator $U(t)=\exp[-it\tilde A]$ directly through in-built Qiskit~\cite{Qiskit} algorithms where we saw no difference in performance. In a practical implementation on a physical quantum device, if one were to construct $U(t)$ through Trotterization, then a smaller number of Pauli strings would benefit from a reduced number of entangling gates.

\begin{figure}[htb]
\begin{center}
\includegraphics[width=0.49\textwidth]{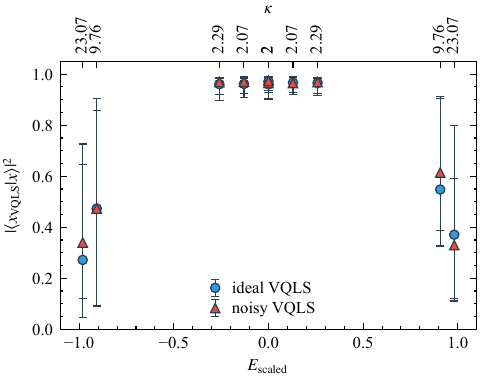}
\caption{\protect  Results from ideal and noisy \texttt{Fake\_Torino} VQLS simulations. The vertical axis is the  overlap $|\bra{x_\text{QVLS}}x\rangle|^2$ of the optimized trial wavefunction $\ket{x_\text{QVLS}}=|\alpha_0\rangle$ with the exact result $|x\rangle$ for the 2-site device. The horizontal axis on the bottom shows the scaled energy $E_\text{scaled}$ and the horizontal axis on top shows the condition number $\kappa$ of the $3 \times 3$ block, without the subspace for $\psi_-$, of the matrix  $\xbar{A}$ in Eq.~(\ref{eq:QLSE}). }
\label{fig:2siteVQLSOverlap}
\end{center}
\end{figure}

We experimented with various optimization methods for the classical component of the VQLS algorithm, finally settling on the Powell optimizer~\cite{Powell_1964} provided in SciPy~\cite{2020SciPy-NMeth}. In Fig.~\ref{fig:2siteVQLSOverlap} we show the overlap of the trial wavefunctions with the exact solution $|\langle x_\text{VQLS}|x\rangle|^2$ at the minimum of the local cost function $C_L$ in  Eq.~(\ref{eq:localcost}), for ideal and noisy simulations. The exact result $\ket{x}$ were obtained with the same parameters $\beta=23^\circ$, $t_{ab}=t_L=\SI{2.7}{\eV}$ used earlier for the HHL calculation. The $\ket{x_\text{VQLS}}=|{\bm\alpha}_0\rangle$ results were obtained from simulations  using  Powell optimization with 1000 shots for each circuit.  The local cost was optimized for minimization from up to 150 evaluations of $C_L$. Note that the dragon solution in the quantum algorithm is independent of $\beta$ and $t_L$. Obtaining the solution to the original equation through $\psi_1=\psi_+\cos\beta$, $\psi_2=\psi_+\sin\beta$ results in the overlap shown in Fig.~\ref{fig:2siteVQLSOverlap}. We get the same overlap for other values of $\beta$ and $t_L$, for example, $\beta= 86^\circ$ and $t_L=\SI{0.4}{\eV}$. The calculations were repeated 100 times at 10 energy values $E$ and scaled as in Eq.~(\ref{eq:escaled}) with $E_\text{max/min}= \pm 2 t_L +\epsilon_L$.  We indicate the condition number of the reduced $3\times 3$ block as well. 

From Fig.~\ref{fig:2siteVQLSOverlap}, we see that both for the ideal and noisy simulations, the overlap $|\langle x_\text{VQLS}\ket{x}|^2$ degrades away from the middle of the transmission energy band $-1\leq E_\text{scaled}\leq 1$. Moreover, near the band edges there is a larger spread in the overlap value. We speculate that this is related to the relatively larger condition numbers of the $3\times3$ matrix spanned by the vector ($1+r, \psi_+, s$) near the energy band edges. The condition number as a function of $E_\text{scaled}$ is shown in  Fig.~\ref{fig:2siteVQLSOverlap}. 
As the condition number gets large, small changes in the initial variational state $\ket{\bm \alpha}$ can lead to large changes in the minimized solution $\ket{{\bm \alpha}_0}$. This  in general also leads to a variational solution far from the exact solution $\ket{x}$ in most of the experiments resulting in a median overlap $|\langle{\bm\alpha}_0\ket{x}|^2=|\langle x_\text{VQLS}\ket{x}|^2\ll1$. The differences between the ideal and noisy circuits are not as significant as those observed in the HHL simulations because the  VQLS circuits are not very deep. The inability to accurately calculate the exact result in VQLS is mostly related with the shortcomings in the classical optimizer in arriving at the correct result due to the complexity of the parameter space as the condition number $\kappa$ gets large. 

Note that, by construction,  this degradation in wavefunction overlap does not propagate into a corresponding error in the transmission coefficient $\mathcal{T}(E)$. The physical constraint in Eq.~(\ref{eq:constraint}) is enforced directly on the variational ansatz, restricting the optimization manifold to states that already satisfy perfect transmission. Consequently and by construction, $\mathcal{T}(E) = 1$ is recovered across the entire conducting bandwidth regardless of the optimizer's wavefunction fidelity. This robustness of the physical observable against ill-conditioning highlights the importance of embedding symmetry constraints into near-term variational quantum algorithms.

\begin{figure}[htb]
\begin{center}
\includegraphics[width=0.49\textwidth]{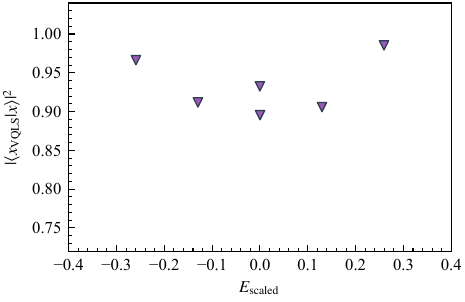}
\caption{\protect  Results from VQLS computations on \texttt{ibmq\_torino} for the 2-site device.  The the  overlap $|\langle \bm\alpha_0|x\rangle|^2$ as a function of scaled energy $E_\text{scaled}$ and condition number $\kappa$. The axis labels are the same from Fig.~\ref{fig:2siteVQLSOverlap}. }
\label{fig:2siteDragonQPU}
\end{center}
\end{figure}

In Fig.~\ref{fig:2siteDragonQPU} we show the results from calculations on physical quantum device \verb|ibm_torino| for  $t_L=\SI{2.7}{\eV}$, $\epsilon_L=\SI{-2.7}{\eV}$ at energies $E=\SI{-4.1}{\eV}$, $\SI{-3.4}{\eV}$,  $\SI{-2.701}{\eV}$,  $\SI{-2.699}{\eV}$,  $\SI{-2}{\eV}$ and  $\SI{-1.3}{\eV}$ corresponding to $E_\text{scaled}=-0.260$, -0.130, $\num{-1.90E-4}$, $\num{1.90E-4}$,  $0.130$ and $0.260$ where the condition number $\kappa$ is not too large, and noisy simulations usually converges to the exact solution within about 150 evaluations of $C_L$ using Powell optimization. These results were obtained with 1000 shots per circuit. We only attempted a single optimization run at select energies as access to QPU is limited and can become prohibitively expensive. In Table~[\ref{tab:QPUoverlapresults}], we show numbers for overlap, optimization iterations, lowest cost function evaluation and its corresponding iteration index.

\begin{table}[htb]
\centering
\caption{\protect VQLS simulation results from \texttt{ibm\_torino}. Iteration number here lists number of evaluations of the local cost $C_L({\bm \alpha})$ by the Powell optimizer before termination. The optimal $\ket{x_\text{VQLS}}=|{\bm \alpha}_0\rangle$ is chosen from the state with the lowest cost.}
\begin{ruledtabular}
\begin{tabular}{cccc}
$E$ & $E_\text{scaled}$ & Iterations & $|\langle x_\text{VQLS}\ket{x}|^2$
\\ \hline \rule{0pt}{0.9\normalbaselineskip}
\begin{filecontents*}{Table_IBM_Torino_VQLS.csv}
energy,escaled,iterations,overlap,te
-4.1,-0.26,90,0.9663,1
-3.4,-0.13,69,0.9118,1
-2.701,-1.90E-04,102,0.9325,1
-2.699,1.90E-04,90,0.8954,1
-2,0.13,141,0.9058,1
-1.3,0.26,148,0.9854,1
\end{filecontents*}
\csvreader[head to column names, late after line=\\]{Table_IBM_Torino_VQLS.csv}{}
{\ \num{\energy} &  \num{\escaled} &\iterations &\overlap}
\end{tabular}
\end{ruledtabular}
 \label{tab:QPUoverlapresults}
\end{table}

\subsection{6-site Quantum Dragon}
\label{subsec:results_6site}

The 6-site device with $L=14$ generates $(1+3)L^2-L=770$ 3-qubit circuits. In the IBMQ machines we stack many 3-qubit circuits to be evaluated in parallel into a layout of available qubits. However, this device in general requires much more cost function evaluations compared to the 2-site device with $L=6$. This together with the larger  number of circuits would require more computation time than available to us. So, for this device we only provide ideal simulation results from Qulacs~\cite{Suzuki_2021}. We perform up to 600 evaluations of the cost function $C_L$ for the Powell optimizer with 1000 shots per circuit in the simulator for each evaluation.  

\begin{figure}[htb]
\begin{center}
\includegraphics[width=0.49\textwidth]{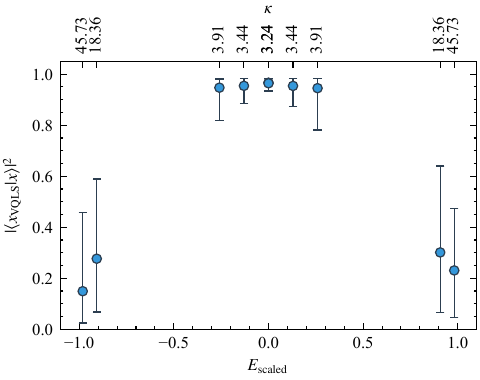}
\caption{\protect Results from ideal simulations for the 6-site device. The the  overlap $|\langle \bm\alpha_0|x\rangle|^2$ as a function of $E_\text{scaled}$ and  $\kappa$.  The axis labels are the same from Fig.~\ref{fig:2siteVQLSOverlap}. }
\label{fig:6siteDragonOverlap}
\end{center}
\end{figure}

In Fig.~\ref{fig:6siteDragonOverlap}, we show the overlap of the trial 
wavefunction at a cost minimum with the exact result. The exact results were for the matrix defined in Eq.~(\ref{Eq:6siteDragon:01}) with  $\beta_j=70^\circ, 50^\circ, 50^\circ $; $t_j=1.9, 0.0, 1.2$ for $j=1,2,3$  and $\kappa_n=0.1, -0.39$ for $n=1,2$, respectively. We see that the overlap is in general not as a great compared to the 2-site device. The increase in the size of the Hilbert space where the optimization has to be performed significantly affects the result. 

\subsection{HHL and VQLS comparison}
\label{subsec:HHLSandVQLS}

A qualitative and quantitative comparison of the HHL and VQLS algorithms for the dragon solutions can be made from the 2- and 6-site device calculations. 
From the 2-site device calculations in  Figs.~\ref{fig:HHLoverlap}, \ref{fig:HHLtransmission}, \ref{fig:2siteVQLSOverlap} and ~\ref{fig:2siteDragonQPU}, we see HHL can be used for nanodevice calculations. The algorithmic overhead includes an ancilla qubit for block encoding non-Hermitian matrix $A$ (or $\xbar{A})$, and several clock qubits for spectral decomposition of  $A$. On NISQ quantum devices this is not a major strain that currently boast hundreds of qubits. The major drawback is the deep circuits required for the QPE in the HHL algorithm which involve a large number of entangling gates that are still too noisy for the algorithm to be practical. This is demonstrated in the noisy HHL simulations. 

The VQLS algorithm in contrast to HHL requires fewer qubit registers. It is also less susceptible to noise for two reasons. First, we are able to construct physically motivated variational ans{\"a}tze $\ket{\bm\alpha}$ that are relatively shallower compared to the QPE circuits. Second, though the cost function evaluations involve entangling gates, they are again not as deep as the QPE circuits. The advantage of the finely tuned variational state is that the transmission coefficient is always perfect $\mathcal{T}(E)=1$. The drawback in the variational calculation is the heuristic classical optimizers that are used for cost minimization. The problem of barren plateaus in the classical optimization is known~\cite{larocca2024review}. As the system size grows, see Fig.~\ref{fig:6siteDragonOverlap}, accurate calculation of the ground state $\ket{x}$ dragon solution gets harder. The problem is made worse at energies where the condition number is large. The accuracy cannot be improved by increasing the number of shots in the quantum computation. The HHL results seem to be unaffected by the large condition numbers of the matrices we considered. On the other hand, VQLS does provide an opportunity to perform reliable computations in a narrow energy band where the condition number is small when paired with a physically motivated ansatz.

During the writing of this manuscript, we learned of a recent work by Yang et al.~\cite{yang2025} that also proposes VQLS to simulate quantum transport in nanodevices. However, their work was not on quantum dragon devices where $100$\% transmission is required.

\section{Conclusions}
\label{sec:conclusions}

We have presented the first quantum-computer implementation of the NEGF formalism for nano-scale electron transport, applying both the HHL and VQLS algorithms to compute the electron transport coefficient $\mathcal{T}(E)$ at scattering energy $E$ for dragon nanodevices within the single-band tight-binding model. The problem maps onto compact circuits of $\log_2(2+\ell m)$ system qubits—2 and 3 physical qubits for the 2-site ($\ell=1, m=2$) and 6-site ($\ell=3, m=2$) devices, respectively---while similarity transformations that block-diagonalize the NEGF linear system reduce the number of Pauli strings in the decomposition of the block-encoded matrix, shortening the number of expectation-value circuits.

The HHL algorithm produces a solution that has a $\gtrsim 97\%$ overlap with the exact dragon state $\ket{x}$ for the 2-site device under ideal simulation with zero noise and 1000 shots. We find a near perfect $\mathcal{T}(E)\approx1$ on simulations with $10^3$ and $10^4$ shots that were repeated 100 times at 10 different energies $E$. The post-selection success probability $P_\text{success}$, which is integral to HHL, matches the expected theoretical value calculated from direct classical numerical method using the statevector construct in Qiskit~\cite{Qiskit}. In the allowed scattering energy band, the $P_\text{success}$ determines what fraction of the measurements in the quantum simulation results in an acceptable solution, and hence controls the efficiency and precision of the result, was about $\sim20\%$. In the noisy simulation, the accuracy of the HHL results drops for the quantum dragon state. The deep QPE circuits inherent in the HHL algorithm reduced the overlap with the exact ground state  around $\sim40\%$ across the energy band which is explained as originating from a very noisy signal where the amplitudes of the wavefunctions are generated randomly with equal probabilities without any structure. 

The 2-site nanodevice has been studied using the quantum-classical hybrid VQLS as well. We have used a physically motivated variational ansatz state for the classical optimization for the ground state search. The algorithm has been found to be robust against noise but variational ground state showed significant sensitivity to the initial trial state as the condition number of the  matrix defining the linear system of equations for the dragon state became large towards the edges of the energy band. The accuracy of the HHL algorithm, in contrast, was unaffected by the variation in the condition number across the energy band. By construction, the variational ansatz imposed the perfect $\mathcal{T}(E)=1$ condition and  we were able to find an energy window where the overlap with the exact solution was more than 90\% in simulation on physical quantum device.

For the 6-site hexagonal quantum dragon---sharing the connectivity relevant to graphene and armchair carbon-nanotube junctions---ideal VQLS simulations with Qulacs~\cite{Suzuki_2021} and up to 600 Powell evaluations gave a lower mean overlap and greater sensitivity to initial conditions compared to the 2-site device, consistent with the harder optimization landscape arising from the larger Hilbert space and higher condition numbers ($\kappa > 45$ near the band edges). However,  we were only able to perform  simulations, ideal and noisy, on classical computers within the resources available to us.

Taken together, these results establish a clear algorithmic division: VQLS is the viable near-term algorithm for NEGF-based transport on NISQ hardware, while HHL is the natural fault-tolerant-era successor whose exponential advantage in system size becomes practically accessible once logical error rates are sufficiently suppressed. Several directions merit further investigation. On the variational side, physics-aware optimizers that explicitly incorporate the current-conservation constraint in Eq.~(\ref{eq:constraint}) into the optimization landscape ---rather than enforcing it only on the ansatz ---may improve convergence at ill-conditioned energies and reduce sensitivity to initial parameters. Scalable ans\"{a}tze that more directly exploit the block structure revealed by the dragon similarity transformation, rather than generic SVD-based preparation, could reduce the effective parameter count and mitigate barren plateaus as the device size grows. For HHL, error-mitigation strategies targeting the dominant noise sources should be pursued. Extensions to devices with more disorder channels, inter-slice connectivity, and realistic many-body corrections, including full tight-binding models of carbon nanotube junctions and DNA molecular wires, constitute the natural next step toward quantum simulation of experimentally relevant nanodevices.

\acknowledgments
This work was partially supported 
 by U.S. DOE Grant No.  DE-SC0024286.
 We acknowledge the use of IBM Quantum services for this work. In
this paper we used \verb|ibmq_torino|.

\appendix

\section{Explicit Values of the Adjustable Dragon Parameters}
\label{app:dragon_params}

\begin{figure}[htb]
\begin{center}
\includegraphics[width=0.52\textwidth,clip=true]{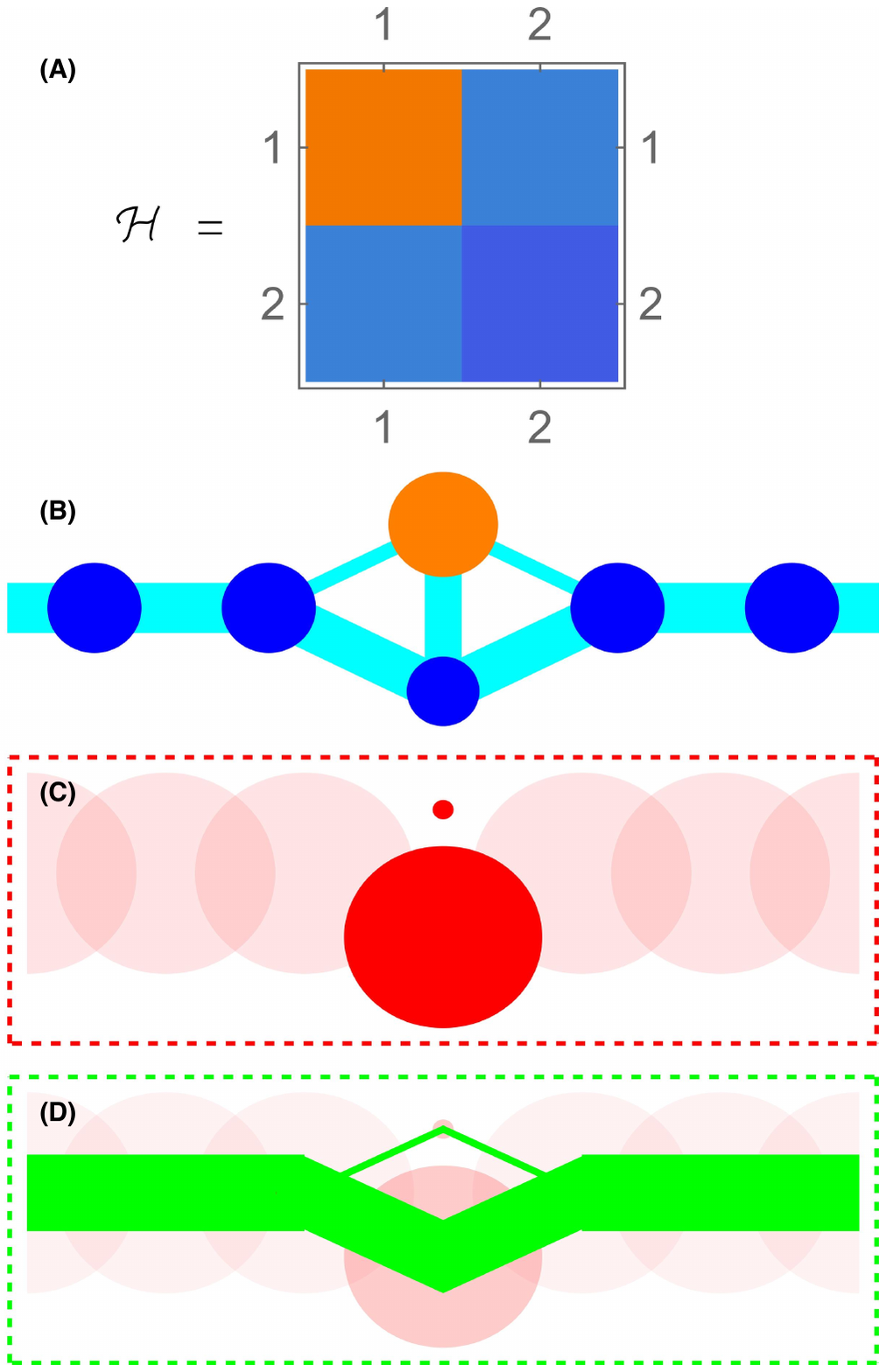}
\end{center}
\caption{Visualization of the 2-site quantum dragon ($\ell=1$, $m=2$) 
nanodevice 
with $\beta=72^\circ$ and $t_{ab}=1.9$~eV and $\delta=0$.  
See the text for a complete description.  
(A) is a plot of the Hamiltonian, 
(B) is a graphic representation of the quantum dragon nanodevice,
(C) is the associated LDOS,
and 
(D) is the associated 
bond currents superimposed on the LDOS.  
\label{fig:2siteDragonAppA}
}
\end{figure}
\begin{figure}[htb]
\begin{center}
\includegraphics[width=0.52\textwidth,clip=true]{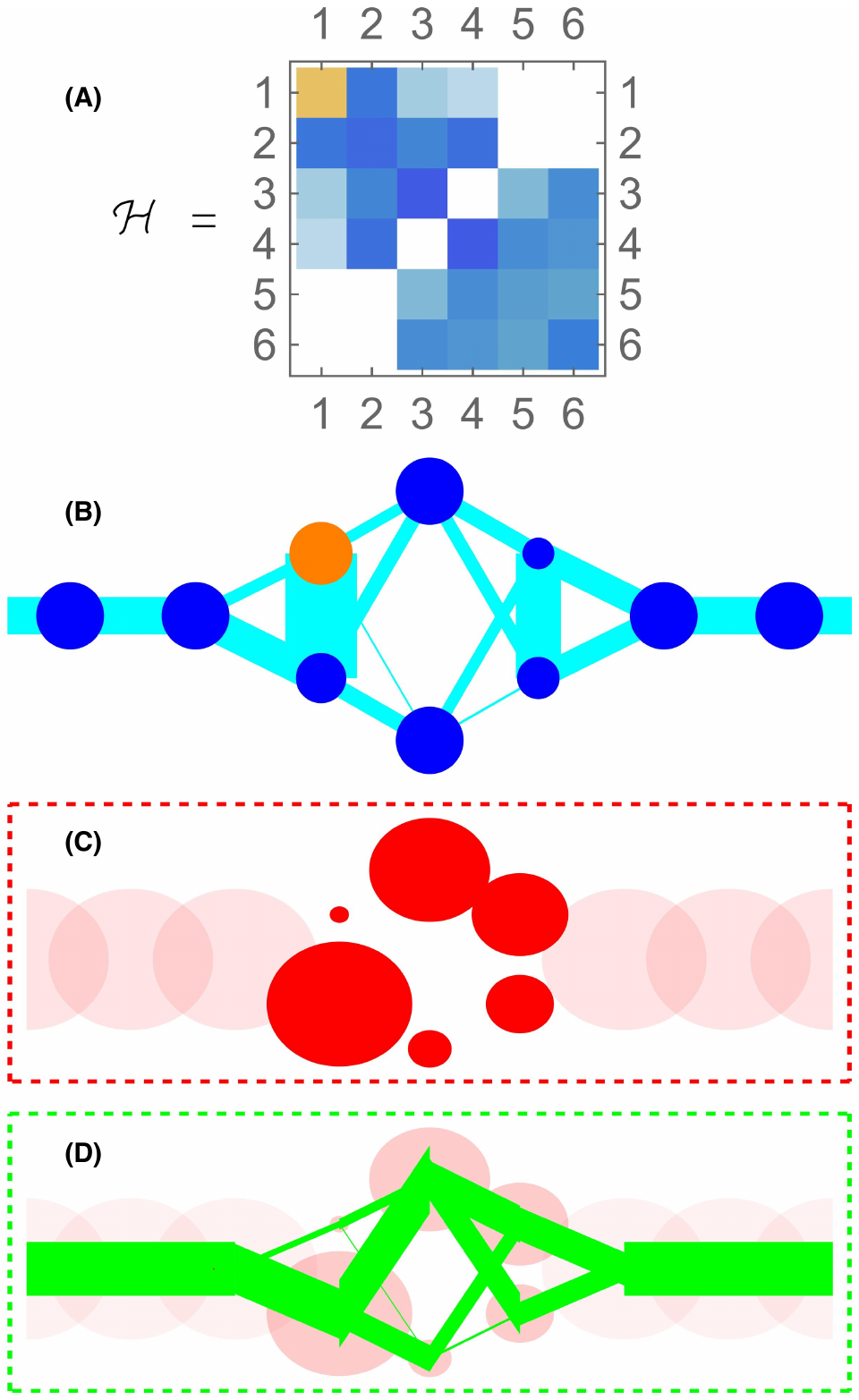}
\end{center}
\caption{\protect 
Visualization of the 6-site quantum dragon nanodevice ($\ell=3$, $m=2$) 
studied using the quantum computer, 
for $\delta$$=$$0$.  
See the text for the complete description.  
Note in (D) all bond currents flow from the left toward the right.
\label{fig:QCdevice6siteAppA}
}
\end{figure}
\begin{figure}[htb]
\centering
\includegraphics[width=0.48\textwidth]{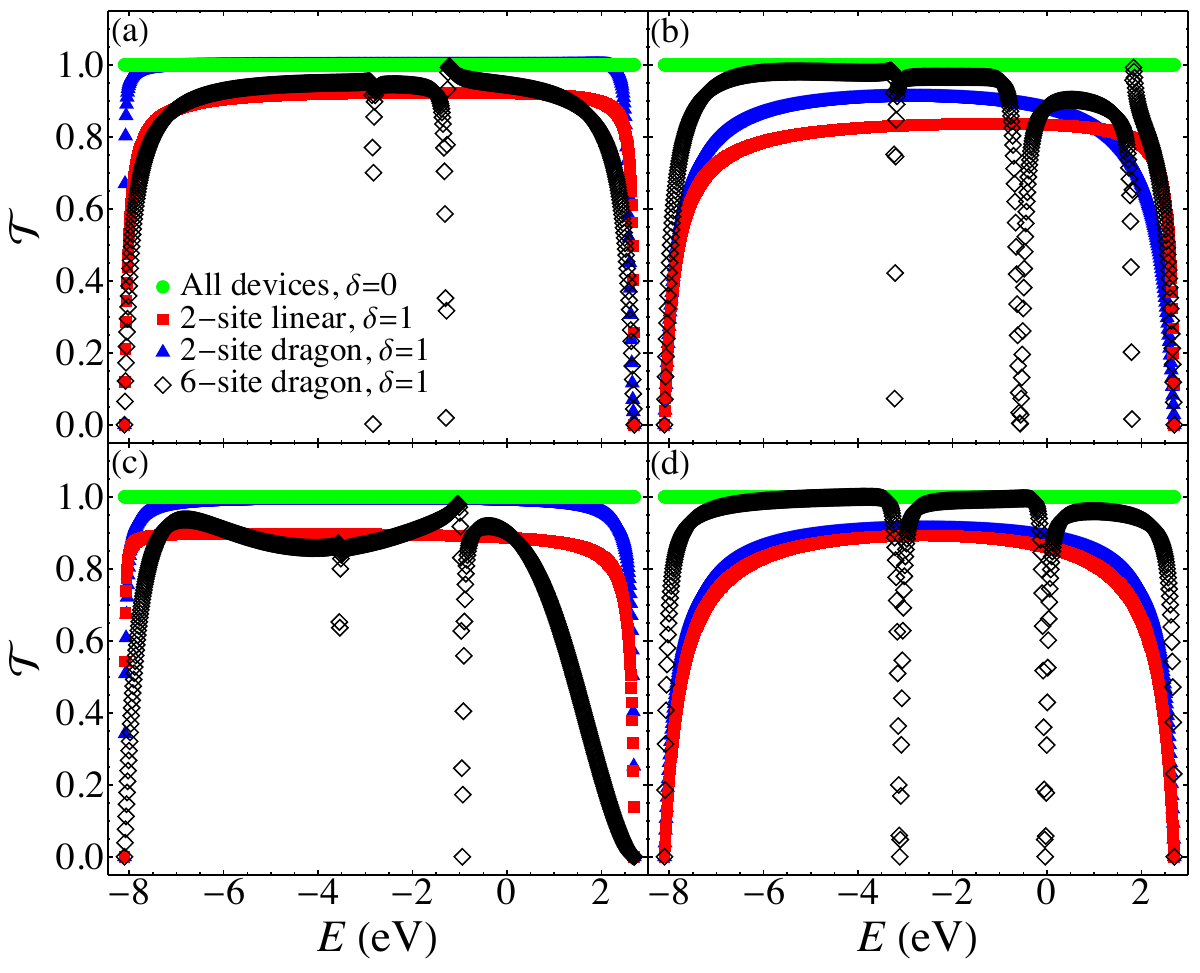}
\caption{\protect 
The transmission ${\cal T}(E)$ is shown for the two 2~site devices 
and the 6~site device.  For each nanodevice $10^3$ energies $E$ were 
calculated.  For all devices when $\delta=0$ for all 
$-8.1\>{\rm ev}<E<2.7~{\rm eV}$ the transmission ${\cal T}(E)=1.0$ 
as shown by the (green) circles on the horizontal line.  
For added quenched disorder $\delta=\SI{1}{\eV}$ the 
2-site device of Sec~\ref{subsec:2siteDragon} are shown as (blue) triangles, 
the 2-site linear device of Sec.~\ref{subsec:2siteslinear} as (red) squares, 
and the 6-site device of Sec.~\ref{subsec:6siteDragon} as (black) open diamonds. The four panels correspond to the four sets of random disorders for the 2-site and 6-site devices mentioned in the text.
}
\label{fig:FanoAntiResonances}
\end{figure}

In this appendix we provide the explicit values, and visualizations, 
of both the 2-site and 6-site quantum dragon nanodevices studied on the 
quantum computer, respectively, Fig.~\ref{fig:2siteDragonAppA}(B) and 
Fig.~\ref{fig:QCdevice6siteAppA}(B).  
The lead parameters are $t_L$$=$$2.7$~eV and $\epsilon_L$$=$$-2.7$~eV as motivated in Sec.~\ref{subsec:results_2site}, so a model of a single 
walled armchair carbon nanotube would have on site energy zero.  
For these nanodevices only the last two atoms of the semi-infinite 
incoming lead and 
first two atoms of the semi-infinite outgoing lead are shown.  
The radii of the (cyan) bonds are proportional to the hopping strengths.  
The radii of the disks are proportional to the absolute value of 
the on site energy, with orange (blue) for positive (negative) values.  
Even though both nanodevice Hamiltonians are disordered, 
each device satisfies the quantum dragon condition of Eq.~\eqref{eq:dragon_cond_B}, yielding 
$\mathcal{T}(E)$$=$$1$ across the full conducting band $\epsilon_L - 2t_L \leq E \leq \epsilon_L + 2t_L$ defined in Eq.~\eqref{eq:propagating}.  
This full transmission for all $-8.1\>{\rm eV}<E<2.7\>{\rm eV}$ is shown in 
Fig.~\ref{fig:FanoAntiResonances} as the horizontal green line.  

For the 2-site device of Fig.~\ref{fig:2sitesDragon} 
we choose $t_{ab}=1.9$~eV and $\beta=72^{\rm o}$.  
The explicit 
$2$$\times$$2$ Hamiltonian as in Eq.~(\ref{eq:2siteNonLinearDevice}) 
is thus
\begin{equation}
\label{Eq:AppA:2siteH}
\begin{array}{lcl}
{\cal H} & = &
\left(\begin{array}{cc}
3.1478\>{\rm eV} & -1.9 \> {\rm eV} \\
-1.9 \> {\rm eV} & -2.0827 \> {\rm eV}
\end{array}\right)
\\
& & \quad + 
\delta
\left(\begin{array}{cc}
-1.1588 & 0 \\
 0 & 0.4091
\end{array}\right)
\>.
\end{array}
\end{equation}
The Hamiltonian of Eq.~(\ref{Eq:AppA:2siteH}) 
with $\delta=0$ is also shown graphically in 
Fig.~\ref{fig:2siteDragonAppA}(A), 
via the {\tt Mathematica} function {\tt MatrixPlot}.  
The first matrix on the right hand side (RHS) of Eq.~(\ref{Eq:AppA:2siteH}) 
is the quantum dragon nanodevice Hamiltonian.  
The second matrix on the RHS of Eq.~(\ref{Eq:AppA:2siteH}) 
is random quenched on site disorder chosen from a 
Gaussian distribution of width unity and mean zero $\mathcal N(1,0)$, and 
is multiplied by a strength $\delta$ for this additional 
uncorrelated disorder.

The transmission ${\cal T}(E)$ for this 2-site 
device  for 10$^3$ equally spaced energies 
between $-8.1$~eV and $2.7$~eV is shown in 
Fig.~\ref{fig:FanoAntiResonances} (a) as the 
green symbols on a horizontal line for $\delta=0$ 
and as the blue symbols for $\delta=1$~eV.  
By forming in the normal NEGF manner the retarded Green's function 
${\cal G}$ one obtains the Local Density of States (LDOS) and 
the bond currents. The Green's function, and hence the 
LDOS and bond currents are calculated at 
a particular energy.  However, for visualization purposes 
these are rescaled by a constant value, and for quantum 
dragons these rescaled values for $\delta=0$ 
are independent of 
energy \cite{Novotny_2023}.  
The 2~site LDOS with $\delta=0$ is shown in 
Fig.~\ref{fig:2siteDragonAppA}(C), with the radius of 
the red disks proportional to the LDOS on that site.  
For visual benefit, the LDOS of the lead atoms are shown 
in a lighter shade of red.  
The bond currents with $\delta=0$ are shown 
as the green trapezoids in Fig.~\ref{fig:2siteDragonAppA}(D), 
where the width of the trapezoids are proportional to the 
current flowing to the right in that bond.  The currents 
are superimposed on the red disks representing the LDOS.  
Note because ${\cal T}(E)$$=$$1$ the width of the rectangles 
in the incoming and outgoing leads are identical.  In Fig.~\ref{fig:FanoAntiResonances}, we illustrate results for  three additional random disorder drawn from $\mathcal N(1,0)$: $\operatorname{diag}(0.5370 , -1.8559)$, $\operatorname{diag}(1.1963, -0.6261)$, $\operatorname{diag}(0.0417, 1.9185)$.

Although not studied on the quantum computer, for comparison we calculated the transmission for the 2-site linear chain of Sec.~\ref{subsec:2siteslinear}.
For $t_{ab}$$=$$2.7$~eV we obtain ${\cal T}(E)$$=$$1$, the 
green symbols on a  horizontal line in Fig.~\ref{fig:FanoAntiResonances}.  
We also include the same 4 sets of quenched random disorder as the 2-site dragon device and show in Fig.~\ref{fig:FanoAntiResonances} 
the ${\cal T}(E)$ as red symbols.  

For the 6~site dragon ($m$$=$$2$ and $\ell$$=$$3$) of Sec.~\ref{subsec:6siteDragon} and 
Fig.~\ref{fig:6sitesDragon}
we choose the three intra-slice hopping 
$t_{j,{\rm intra}}$ to be $1.9,\> 0.0,\>1.2$ for $j=1,2,3$.  
The quantum dragon nanodevice is shown in 
Fig.~\ref{fig:QCdevice6siteAppA}(B).  Because we have chosen 
zero hopping in the middle slice, we draw the 6~slice quantum dragon 
as a hexagon (analogous to a benzene ring).  
We chose the parameters $\beta_j = 70^\circ, 50^\circ, 50^\circ$.  
We also choose the inter-slice singular values $\kappa_n = 0.1, -0.39$, satisfying the bounds of Eq.~\eqref{Eq:6site:General:05}
The description for the 6~site quantum dragon of 
Fig.~\ref{fig:QCdevice6siteAppA} is the same as 
that for the 2~site quantum dragon of Fig.~\ref{fig:2siteDragonAppA}.  
In Fig.~\ref{fig:QCdevice6siteAppA} 
panel A shows the $6$$\times$$6$ device Hamiltonian matrix; 
panels B--D show the device geometry, LDOS, and bond currents, respectively. 
The visual code for the nanodevice, LDOS, and 
bond current are the same as used in 
Fig.~\ref{fig:2siteDragonAppA}.  
Even for this strongly disordered nanodevice 
the transmission ${\cal T}(E)$$=$$1$ 
for all electron energies that propagate 
through the leads, as shown 
by the green horizontal lines in Fig.~\ref{fig:FanoAntiResonances}, so 
it is a quantum dragon nanodevice.  
Note the quantum dragon in 
Fig.~\ref{fig:QCdevice6siteAppA} has 
a different $\vec{v}_\text{dragon}$ for 
each slice it does not show 
order-amidst-disorder 
\cite{Novotny_2021,Novotny_2023} as 
the LDOS on every device site is different.  
Also again note that because ${\cal T}(E)=1$ 
the width of the rectangles in the incoming and 
outgoing leads are identical in panel~D.

Fig.~\ref{fig:FanoAntiResonances} shows the 
transmission ${\cal T}(E)=1$ for $\delta=0$ as the 
green symbols on a horizontal line, 
and as the black symbols for $\delta=\SI{1}{\eV}$, for the 6-site device.  
In the four panels in 
Fig.~\ref{fig:FanoAntiResonances}, we used the following random on site disorder drawn from 
$\mathcal N(1,0)$ :  
$\operatorname{diag}(-0.4068,  0.0176,  0.1011, -0.1998, -0.8795,\allowbreak -1.4964)$, $\operatorname{diag}(-1.4841, -0.8752, -0.6242, -0.2099, \allowbreak -0.9895, \allowbreak 0.501)$, $\operatorname{diag}(-0.2007, -2.1649, -0.872 , -0.66  , \allowbreak-0.4506, -1.1456)$,
$\allowbreak\operatorname{diag}(0.1022, -0.594 , -0.7823, 
0.2066,  \allowbreak 0.5578, -0.3387)$.
It has recently been shown for on site 
energy disorder that averaging over 
$\mathcal N(1,0)$ times $\delta$ for 
small $\delta$ gives universal scaling 
in two different regimes for the 
average transmission ${\cal T}_{\rm aveage}(E)$ 
\cite{Novotny2025} for all nanodevices 
with ballistic electron propagation  
and for all quantum dragon nanodevices.  
For particular random 
variants from $\mathcal N(1,0)$ times 
strength $\delta$, as seen in 
the four panels in Fig.~\ref{fig:FanoAntiResonances},
the transmission exhibits 
Fano anti-resonances where the transmission function 
goes from near unity to exactly zero and back 
to near unity over some 
small range of energies $E$ \cite{MiroFano2010,Novotny2025}.
For small $\delta$ where these 
anti-resonances occur are very close 
to the eigenvalues of the $\delta$$=$$0$
device Hamiltonian \cite{Novotny2025}.
For large $\delta$ a diffusive 
regime (or for large devices an 
Anderson localization regime) occurs 
with ${\cal T}(E)$$\ll$$1$ with 
probability one for a randomly chosen 
$E$.

%

\section{Tuning the time-evolution parameter $t$ for the dragon solution}
\label{appendix:tuning_t}

A core part of the HHL algorithm is the spectral decomposition of the Hermitian matrix $\tilde A$ in Eq.~(\ref{eq:Atilde}) using QPE. Specifically, the unitary operator $U(t)=\exp(i \tilde{A} t)$ acting on an eigenstate $|\lambda_k\rangle$ of $\tilde A$ is expressed in QPE with a phase 
\begin{align}
    U\ket{\lambda_k}&=e^{i2\pi\varphi_k}\ket{\lambda_k}& \implies
    \varphi_k &= \frac{\lambda_k\, t}{2\pi} \in [0,1)\,. 
\end{align}
The phase $\varphi_k$ is represented in the $n_c$ clock qubit register to a precision of $\delta\varphi=2^{-n_c}$. Thus, the choice of $t$ affects  how the physical eigenvalue
spectrum is mapped by \verb|ExactReciprocalGate()| in Qiskit~\cite{Qiskit} into the representable phase window, and its choice governs both the accuracy and the circuit-depth requirements of the algorithm. The gap in the spectrum plays a role. 
To distinguish non-degenerate eigenvalues $\lambda_i$ and $\lambda_j$, we need a phase resolution 
\begin{align}
  \frac{|\lambda_i-\lambda_j|t}{2\pi}> 2^{-n_c}\,.   
\end{align}
Thus, at a fixed $n_c$ reducing $t$ below the threshold implied above causes the eigenvalues to collapse into a single clock bin, making $1/\lambda_k$ indeterminate.
The phase encoding also requires $\lambda_k t<2\pi$ to prevent phase aliasing that implies $t<2\pi/\lambda_\text{max}$ where $\lambda_\text{max}$ is the spectral norm of $\tilde A$.   These aspects of the HHL algorithm are well known~\cite{harrow2009}. However, without knowing the spectrum $\lambda_k$ a priori, one cannot efficiently choose HHL parameters $n_c$ and $t$. 

\begin{figure}[htb]
\begin{center}
\includegraphics[width=0.49\textwidth]{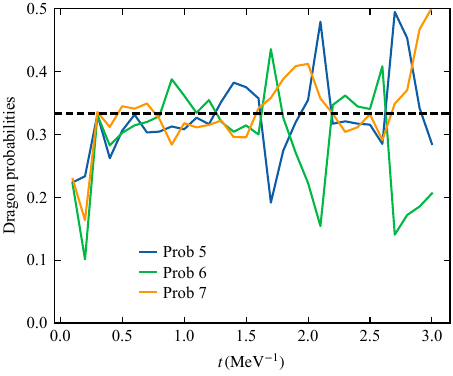}
\caption{\protect Tuning $t$ by looking at the expected nonzero probabilities of the 8-dimensional probability measurements in the $n_b=3$ block-qubit registers. Results shown at $E=\SI{-8}{\eV}$ with a statevector calculation. Details are in the text.}
\label{fig:tunning_t}
\end{center}
\end{figure}

\begin{figure}[htb]
\begin{center}
\includegraphics[width=0.49\textwidth]{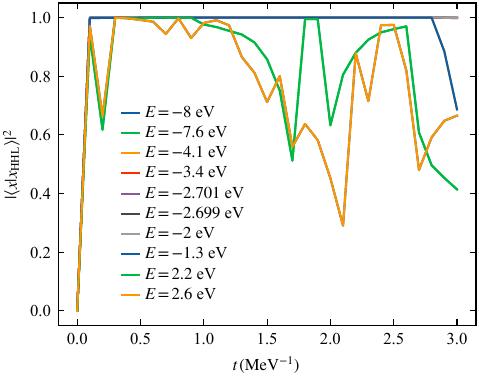}
\caption{\protect Statevector calculation overlap $|\langle{\tilde x}|x_\text{HHL}\rangle|^2$ as a function of time-evolution parameter $t$ at various energies considered in the main text. 
}
\label{fig:2siteDeltaOverlap}
\end{center}
\end{figure}

For the problem we wish to solve of finding the dragon solution of perfect transmission in nonlinear nanodevices, we can tune $t$  such that the probabilities measured in the physical block-qubit registers $n_b$ has the desired structure. For the 2-site device that is encoded with $n_b=3$, we can tune $t$ such that of the 8 probability measurements, the \qtyrange{5}{7}{th} probabilities are all nearly 1/3. For the 6-site device where $n_b=4$, we would tune $t$ such that the \qtyrange{9}{13}{th} out of the 16 probabilities are each $1/5$.  In Fig.~\ref{fig:tunning_t}, we show the \qtyrange{5}{7}{th} measured probabilities in a statevector calculation as a function of $t$ at energy $E=\SI{-8}{\eV}$ with $n_c=7$, scale factor $s'=0.5$ in \verb|ExactReciprocalGate()|. We can pick, for example, $t=0.3$ where all the three probabilities are nearly equal to 1/3. For large LSE where classical statevector calculations are not feasible, one would have to perform the expensive tuning of $t$ on physical quantum computers. In Fig.~\ref{fig:2siteDeltaOverlap}, we show the statevector overlap $|\langle{\tilde x}|x_\text{HHL}\rangle|^2$ for a range of $t$ values. One can see that for the $t$ values suggested by Fig.~\ref{fig:tunning_t} for equal probabilities leads to $|\langle{\tilde x}|x_\text{HHL}\rangle|^2\approx 1$ for $E=\SI{-8}{\eV}$.  

\input{QuantumDragon.bbl}
\end{document}

%% file: QuantumDragon.bbl
%